\pdfoutput=1
\documentclass[letterpaper,11pt]{article}
\usepackage{amsmath,amssymb,amsfonts}
\usepackage[multiple,stable]{footmisc}
\allowdisplaybreaks[4]
\usepackage[noconfig]{refstyle}
\usepackage[dvipsnames]{xcolor}
\usepackage[hyperfootnotes=false, linktocpage=true, colorlinks, citecolor=blue, linkcolor=blue, urlcolor=Maroon]{hyperref}
\usepackage{array,tabularx,booktabs,subfigure,graphicx,longtable}
\usepackage[bf]{caption}
\usepackage[paper=letterpaper,margin=1in]{geometry}
\usepackage{tikz}
\usetikzlibrary{shapes,matrix,arrows}
\usepackage{url,lineno}
\usepackage{bm}
\usepackage{cancel}
\usepackage{enumitem}
\usepackage{blkarray}
\usepackage{multirow}
\usepackage{hhline}
\usepackage{listings}
\usepackage{verbatim}
\usepackage{footmisc}
\usepackage{tensor}
\usepackage{bbm}
\usepackage{mathtools}
\usepackage{graphicx}
\usepackage{listings}
\usepackage{caption}
\usepackage{tikz}
\usetikzlibrary{trees}
\usepackage{cite}
\usepackage{blkarray}
\usepackage{float}

\usepackage{listings}
\usepackage{hyperref}
\usepackage{xcolor}

\definecolor{bgd}{rgb}{.88,.87,.88}
\definecolor{sage}{rgb}{.35,.21,.85}
\usepackage{appendix}

\usepackage{indentfirst}

\tikzstyle{block} = [rectangle, draw, text width=7em, text centered, rounded corners, minimum height=3em]

\newcommand{\iddots}{\mathinner{\mkern2mu\raise1pt\hbox{.}\mkern2mu \raise4pt\hbox{.}\mkern2mu\raise7pt\hbox{.}\mkern1mu}}

\newcommand{\cV}{\mathcal{V}}
\newcommand{\hatcV}{\hat{\mathcal{V}}}
\newcommand{\hatt}{\hat{t}}
\newcommand{\hattau}{\hat{\tau}}
\newcommand{\icV}{\left(\mathcal{V}^{-1}\right)}
\newcommand{\ihatcV}{\left(\hat{\mathcal{V}}^{-1}\right)}
\newcommand{\cK}{\mathcal{K}}
\newcommand{\icK}{\left(\mathcal{K}^{-1}\right)}

\newcommand{\icKK}{\left(\mathcal{K}_{T}^{-1}\right)}
\newcommand{\der}[2]{\frac{\partial #1}{\partial #2}}
\newcommand{\derder}[3]{\frac{\partial^{2} #1}{\partial #2 \partial #3}}

\providecommand{\id}{\leavevmode\hbox{\small$\mathrm{1}$\kern-3.8pt\normalsize$\mathrm{1}$}}
\def\fnote#1#2{\begingroup\def\thefootnote{#1}\footnote{#2}
     \addtocounter{footnote}{-1}\endgroup}

\def\sqz{.5}
\def\sqzinverse{2}
\def\hsqz{1}
\newlength\bls
\newlength\tmplength
\def\scalebrace#1#2{\tmplength=#1\bls\relax%
  \scalebox{\hsqz}[\sqz]{\rotatebox{90}{$\underbrace{\hspace{\sqzinverse\tmplength}}$}}%
  \raisebox{\dimexpr+.5\tmplength+.5\dp\strutbox-.5\ht\strutbox}{$\scriptstyle\;\; #2$}}

\begin{document}

\vspace{1cm}

\title{
       \vskip 40pt
\textbf{      {\Huge New Large Volume Solutions}}}
      \vspace{2cm}

\author{\textbf{Ross Altman${}^a$, Yang-Hui He${}^{b}$, Vishnu Jejjala${}^{c}$, and Brent D.\ Nelson${}^{a}$}}
\date{}
\maketitle
\begin{center}${}^a${\small  Department of Physics, Northeastern University, Boston, MA 02115, USA \\
${}^b$ Department of Mathematics, City, University of London, Northampton Square, London EC1V 0HB, UK;
School of Physics, NanKai University, Tianjin, 300071, P.R.\ China; \\
and Merton College, University of Oxford, OX1 4JD, UK \\
${}^c$ Mandelstam Institute for Theoretical Physics, National Institute for Theoretical Physics, CoE-MASS, and School of Physics, University of the Witwatersrand, Johannesburg, WITS 2050, South Africa} \\
\fnote{}{${}$\hspace{-0.25in}altman.ro@husky.neu.edu, hey@maths.ox.ac.uk, vishnu@neo.phys.wits.ac.za, b.nelson@neu.edu}
\end{center}

\begin{abstract}
In previous work, we have commenced the task of unpacking the $473,800,776$ reflexive polyhedra by Kreuzer and Skarke into a database of Calabi--Yau threefolds~\cite{Altman:2014bfa} (see~\url{www.rossealtman.com}).
In this paper, following a pedagogical introduction, we present a new algorithm to isolate Swiss cheese solutions characterized by ``holes,'' or small 4-cycles, descending from the toric divisors inherent to the original four dimensional reflexive polyhedra.
Implementing these methods, we find $2,268$ explicit Swiss cheese manifolds, over half of which have $h^{1,1}=6$.
Many of our solutions have multiple large cycles.
Such Swiss cheese geometries facilitate moduli stabilization in string compactifications and provide flat directions for cosmological inflation.
\end{abstract}

\thispagestyle{empty}
\setcounter{page}{0}
\newpage

\setcounter{tocdepth}{2}
\tableofcontents

\numberwithin{equation}{section}
\def\theequation{\thesection.\arabic{equation}}

\renewcommand\tilde{\widetilde}

\section{Introduction}\label{intro}

Kreuzer and Skarke have exhaustively classified the $473,800,776$ reflexive polyhedra in four dimensions~\cite{Kreuzer:2000xy}.
Each of these reflexive polyhedra gives rise to a four dimensional toric variety in which the anticanonical hypersurface is a singular Calabi--Yau threefold~\cite{Batyrev:1994hm}.
Moreover, each of these singular hypersurfaces admits at least one, but potentially many \textit{maximal projective crepant partial} (MPCP) desingularizations, some of which represent adjacent regions in the moduli space of the same manifold, and some of which are entirely independent.
This leaves us with an indeterminate, but undeniably large class of Calabi--Yau threefolds, well in excess of the half billion reflexive polytopes.
In a previous work~\cite{Altman:2014bfa}, we have started to compile a catalog of Calabi--Yau threefolds extracted from the Kreuzer--Skarke dataset~\cite{Kreuzera} into a new database indexed by the topological and geometric properties of the threefolds (see \url{www.rossealtman.com}~\cite{Altmana}).
As important features of geometries for compactification are readily available in a format that can be queried or scanned in batch, our database provides an efficient and useful resource for the string phenomenology and string cosmology communities.

With the enormous number of candidate Calabi--Yau compactifications in hand, model builders are confronted with the challenge of isolating the set of constructions which might potentially replicate physics in the real world.
In type IIB string theory, the particularly difficult problem of moduli stabilization can be avoided via flux considerations in one of two prevailing Calabi--Yau threefold compactification paradigms: KKLT~\cite{kklt} or the large volume scenario~\cite{Bobkov:2004cy, bbcq, Cicoli:2008va}.
A particularly interesting subset of the latter are the so-called ``Swiss cheese'' compactifications.
The name derives from the fact that a subset of the K\"ahler moduli are large and control the overall volume of the manifold, while the the rest of the K\"ahler moduli remain small and control the volumes of the ``holes'' at which non-perturbative contributions to the superpotential, such as E$3$-instantons, are localized.
In this paper, we consider a special subclass of Swiss cheese compactifications characterized by large and small cycles that descend directly from the toric divisors of the Calabi--Yau threefold and are therefore directly encoded in the four dimensional reflexive polyhedra of Kreuzer and Skarke.
We detail an algorithm for identifying such geometries.

Implementing this algorithm, we have conducted a first scan of the current database of Calabi--Yau threefolds ($h^{1,1}\leq 6$) for the existence of the special class of Swiss cheese geometries, which we refer to as the \textit{toric Swiss cheese} solutions.
When we find a solution of this type, we compute the rotation matrices from the given basis of 2-cycle and 4-cycle volumes (represented by $t^{i}$ and $\tau_{i}$, respectively) into the bases where the large and small cycles are manifest.
Our main result is to report the data for $2,268$ of these toric Swiss cheese Calabi--Yau geometries, over half of which have $h^{1,1}=6$.
Of these, $70$ have two or more large cycles.
The full details are available in the database of toric Calabi--Yau threefolds located at \url{www.rossealtman.com}.
The number of large cycles in these geometries range from $1$ to $h^{1,1}(X)-1$.


The organization of the paper is as follows.
In Section~\ref{sec:methods}, we outline the conditions for the possible existence of a large volume solution in the language of toric geometry.
This allows us to set our notation and conventions.
The conditions in the general Swiss cheese case are summarized in Subsection~\ref{sec:genconds}, while the particular case of toric Swiss cheese is presented in Subsection~\ref{toriccase}.
Section~\ref{sec:algorithm} contains a schematic of the algorithm used in \texttt{Sage} to detect and compute toric Swiss cheese solutions for the various Calabi--Yau threefold vacua.
In preparation for presenting our results, we establish in Section~\ref{sec:class} some terminology on classifying large volume solutions on the basis of the form of the Calabi--Yau volume.
As terminology is used by various groups in slightly different contexts, we hope that this classification helps disambiguate language often used in the literature.
In Section~\ref{sec:examples}, we show an explicit example of a Swiss cheese manifold with Hodge numbers $(h^{1,1},h^{2,1}) = (4,94)$ and two large cycles, and perform a minimization of the potential.
We then present our results and discuss some implications in Section~\ref{sec:conc}.
Finally, Appendices ~\ref{sec:appa} and ~\ref{sec:appb} provide a self-contained pedagogical review on the background of the Large Volume Scenario (LVS).

\section{Methods: Detecting Toric Swiss Cheese Solutions}\label{sec:methods}

Consider a Calabi-Yau threefold hypersurface $X$ in an ambient 4-dimensional toric variety $\mathcal{A}$ with $k$ toric coordinates $x_{1},...,x_{k}$, each corresponding to a divisor $D_{i}=\{x_{i}=0\}$ on $X$. The K\"ahler moduli space is given by $H^{1,1}(X)\cap H^{2}(X;\mathbb{Z})$ with dimension $h^{1,1}(X)=\text{dim }H^{1,1}(X)$; we shall be largely concerned with the so-called \textit{favorable} manifolds where all divisor classes on $X$ descend from that of the ambient ${\cal A}$, so that
\begin{equation}
h:=h^{1,1}(X)=h^{1,1}(\mathcal{A})\, .
\end{equation}

A $\mathbb{Z}$-basis of 2-form classes, corresponding to 4-cycles in homology via Poincar\'e duality, can be chosen as $\{J_{1},...,J_{h}\}\in H^{1,1}(X)\cap H^{2}(X;\mathbb{Z})$ spanning the space. Because a Calabi-Yau manifold is K\"ahler, it is naturally equipped with a characteristic K\"ahler 2-form class $J\in H^{1,1}(X)\cap H^{2}(X;\mathbb{Z})$. Expanding the K\"ahler form in a basis, we find

\begin{equation}
J=t^{i}J_{i}\, ,
\end{equation}

\noindent with K\"ahler parameters $t^{i}\in\mathbb{Z}$. However, the K\"ahler form itself is basis independent, and we can therefore choose any basis\footnote{Note: the superscript Latin characters $A, B, C,\dots$ are labels rather than indices, and will \textbf{not} obey the Einstein summation convention.} $\{J^{A}_{1},...,J^{A}_{h}\}\in H^{1,1}(X)\cap H^{2}(X;\mathbb{Q})$, where, for the sake of computational efficiency, we have relaxed the requirement of $\mathbb{Z}$-valued coefficients to the more general case of $\mathbb{Q}$-valued ones. We can then expand the K\"ahler form in this new $A$-basis as follows
\begin{equation}
J=t^{Ai}J^{A}_{i} \, . \label{met:basisexpansion}
\end{equation}

Note that because $J$ is basis-independent, we can easily do this as many times as we want with new bases $B$, $C$, etc. The cohomology or Chow ring structure on $X$, however, \textit{is} basis-dependent. In this chapter, we wish to identify a ``Swiss cheese'' basis in which the large, volume-modulating 4-cycles are manifestly separated from the small, blowup 4-cycles, which are phenomenologically useful in achieving moduli stabilization. But since we have no natural choice of basis to work with, finding one which satisfies the Swiss cheese condition \cite{Gray:2012jy} must involve an arbitrary basis change with many unconstrained degrees of freedom. It is therefore an extremely computationally expensive undertaking, especially when faced with higher dimensional moduli spaces.  Therefore, in order to work around this bottleneck, the only options remaining are to narrow the scope of the search to a special case or to find a particularly natural basis to work with. Later, we will outline a technique that is a combination of these two approaches.

We now consider only the class of smooth toric Calabi-Yau threefolds \cite{Kreuzer:2000xy}, i.e. those obtained as the anticanonical hypersurface in a 4-dimensional toric variety with no worse than terminal singularities. A database\cite{Altman:2014bfa} of these Calabi-Yau threefolds is available through a robust search engine at \url{www.rossealtman.com}. The topological and geometric information for these manifolds is presented in an arbitrary $\mathbb{Z}$-basis $\{J_{1},...,J_{h}\}$.

We can define the $A$-basis of the K\"ahler class as a linear transformation of the original basis ${J_{i}}$. This transformation should be invertible, so we define the transformation matrix $\mathbf{T}^{A}\in GL_{h}(\mathbb{Q})$ by
\begin{equation}
J^{A}_{i}=\left(T^{A}\right)_{i}^{\;\; j}J_{j} \, .\label{met:orig2basis}
\end{equation}
In the same manner, we may introduce matrices $\mathbf{T}^{B}, \mathbf{T}^{C}$, etc. for the $B$-, $C$-, etc. basis representations of the K\"ahler class.

\subsection{Volume, Large Cycle, and Small Cycle Conditions}

The complex subvarieties of $X$ can be written in terms of 2-cycle curves $\mathcal{C}^{i}$, 4-cycle divisors $J_{i}$, and the compact Calabi-Yau 6-cycle $X$. Curves are dual to divisors, and can be expressed in a basis $\mathcal{C}^{1},...,\mathcal{C}^{h}\in\mathcal{M}(\mathcal{A})$ of linear functionals on the space of divisors, $\mathcal{C}^{i}:\,H^{1,1}(\mathcal{A})\rightarrow\mathbb{Z}$, where $\mathcal{M}(\mathcal{A})$ is called the Mori cone, or cone of curves. The K\"ahler class $J$ acts as a calibration 2-form on these $2n$-cycles on $X$, fixing their volumes according to\footnote{We have slightly abused notation by writing $J_{i}$ for both the divisor cohomology class and its Poincar\'e dual in homology.}
\begin{align}
\text{vol}(\mathcal{C}^{i})&=\frac{1}{1!}\int\limits_{\mathcal{C}^{i}}{J}=\frac{1}{1!}\int\limits_{\mathcal{C}^{i}}{t^{j}J_{j}}=t^{j}\delta^{i}_{\;\; j}=t^{i} \, ,\\
\text{vol}(J_{i})&=\frac{1}{2!}\int\limits_{J_{i}}{J\wedge J}=\frac{1}{2!}\int\limits_{X}{J_{i}\wedge t^{j}J_{j}\wedge t^{k}J_{k}}=\frac{1}{2}t^{j}t^{k}\kappa_{ijk}:=\tau_{i} \, ,\\
\text{vol}(X)&=\frac{1}{3!}\int\limits_{X}{J\wedge J\wedge J}=\frac{1}{3!}\int\limits_{X}{t^{i}J_{i}\wedge t^{j}J_{j}\wedge t^{k}J_{k}}=\frac{1}{6}t^{i}t^{j}t^{k}\kappa_{ijk}:=\mathcal{V} \, ,
\end{align}
\noindent where $\kappa_{ijk}=\int_{X}{J_{i}\wedge J_{j}\wedge J_{k}}$ is the triple intersection tensor corresponding to the Chow ring structure of the Calabi-Yau threefold $X$. We can expand the volume $\mathcal{V}$ in complete generality by assuming that each of the three copies of $J$ in the integral is written in a different basis
\begin{align}
\mathcal{V}&=\frac{1}{3!}\int\limits_{X}{J\wedge J\wedge J}=\frac{1}{3!}t^{Ai}t^{Bj}t^{Ck}\int\limits_{X}{J^{A}_{i}\wedge J^{B}_{j}\wedge J^{C}_{k}}\notag\\
&=\frac{1}{3!}t^{Ai}t^{Bj}t^{Ck}\int\limits_{X}{\left(\left(T^{A}\right)_{i}^{\;\; r}J_{r}\right)\wedge \left(\left(T^{B}\right)_{j}^{\;\; s}J_{s}\right)\wedge \left(\left(T^{C}\right)_{k}^{\;\; t}J_{t}\right)}\notag\\
&=\frac{1}{3!}t^{Ai}t^{Bj}t^{Ck}\left(T^{A}\right)_{i}^{\;\; r}\left(T^{B}\right)_{j}^{\;\; s}\left(T^{C}\right)_{k}^{\;\; t}\int\limits_{X}{J_{r}\wedge J_{s}\wedge J_{t}}\notag\\
&=\frac{1}{3!}t^{Ai}t^{Bj}t^{Ck}\left(T^{A}\right)_{i}^{\;\; r}\left(T^{B}\right)_{j}^{\;\; s}\left(T^{C}\right)_{k}^{\;\; t}\kappa_{rst}
\, . \label{met:expandvol}
\end{align}
The volume of each of the 4-cycles $\tau_{i}$ can then be written as the derivative of the total volume with respect to each of the 2-cycle volumes $t^{i}$
\begin{align}
\tau^{A}_{i}&=\frac{d\mathcal{V}}{d t^{Ai}}=\frac{d}{d t^{Ai}}\left[\frac{1}{3!}t^{A'i'}t^{B'j'}t^{C'k'}\int\limits_{X}{J^{A'}_{i'}\wedge J^{B'}_{j'}\wedge J^{C'}_{k'}}\right]\notag\\
&=\frac{1}{2}t^{Bj}t^{Ck}\int\limits_{X}{J^{A}_{i}\wedge J^{B}_{j}\wedge J^{C}_{k}}\notag\\
&=\frac{1}{2}t^{Bj}t^{Ck}\left(T^{A}\right)_{i}^{\;\;r}\left(T^{B}\right)_{j}^{\;\; s}\left(T^{C}\right)_{k}^{\;\; t}\int\limits_{X}{J_{r}\wedge J_{s}\wedge J_{t}}\notag\\
&=\frac{1}{2}t^{Bj}t^{Ck}\left(T^{A}\right)_{i}^{\;\; r}\left(T^{B}\right)_{j}^{\;\; s}\left(T^{C}\right)_{k}^{\;\; t}\kappa_{rst} \, . \label{met:tauA}
\end{align}
In a generic basis $J^{A}_{i}$, the K\"ahler moduli may be arbitrarily large or small. When looking at phenomenological models in the large Volume Scenario (LVS), however, we wish to choose a basis in which some set of cycles can shrink to zero size (i.e. small), while the remaining cycles must be left non-zero (i.e. large). Thus, in the following formulation, the number of large and small cycles will be labeled $N_{L}$ and $N_{S}$, respectively, such that $h=N_{L}+N_{S}$. For compactness of notation and in analogy to computational pseudocode, we define the following index intervals
\begin{align}
I^{\text{Toric}}&=\left[1,k\right]&\text{(Toric divisors)}\label{met:Itoric}\\
I\hspace{0.7cm}&=\left[1,h\right]&\text{(Original basis)}\label{met:I}\\
I^{A}\hspace{0.5cm}&=\left[1,h\right],\;\; I^{A}_{L}=\left[1,N_{L}\right],\text{ and }I^{A}_{S}=\left[N_{L}+1,h\right]&\text{(A-basis)}\label{met:IA}\\
I^{B}\hspace{0.5cm}&=\left[1,h\right],\;\; I^{B}_{L}=\left[1,N_{L}\right],\text{ and }I^{B}_{S}=\left[N_{L}+1,h\right]&\text{(B-basis)}\label{met:IB}
\end{align}
\noindent where $k$ is the total number of toric divisors on the resolved Calabi-Yau threefold\footnote{An $n$-dimensional toric variety $\mathcal{A}$ constructed from an $n$-dimensional reflexive lattice polytope $M$ obeys the short exact sequence
\begin{equation*}
0\rightarrow M\rightarrow\bigoplus\limits_{i=1}^{k}{\mathbb{Z}D_{i}}\rightarrow\text{Pic}(\mathcal{A})\cong H^{1,1}(\mathcal{A})\cap H^{2}(\mathcal{A};\mathbb{Z})\rightarrow 0
\end{equation*}
where the $D_{i}$ are toric divisor classes. Therefore, $k=h^{1,1}(\mathcal{A})+\text{dim}(M)=h^{1,1}(\mathcal{A})+\text{dim}_{\mathbb{C}}(\mathcal{A})$. So, when the codimension 1 hypersurface $X\subset\mathcal{A}$ is favorable, we have $k=h^{1,1}(X)+\text{dim}_{\mathbb{C}}(X)+1$. In the case of a Calabi-Yau threefold, $k=h^{1,1}(X)+4$ specifically.} $X$. We will assume that there is a specific basis $\{J^{A}_{i}\}$ such that 
\begin{equation}
t^{Ai}=
\begin{cases}
{\rm Large}, & i\in I^{A}_{L}\\
{\rm Small}, & i\in I^{A}_{S}\label{met:tALS}
\end{cases}
\end{equation}
\noindent and a specific basis $\{J^{B}_{i}\}$ such that
\begin{equation}
\tau^{B}_{i}=
\begin{cases}
{\rm Large}, & i\in I^{B}_{L}\\
{\rm Small}, & i\in I^{B}_{S}\, , \label{met:tauALS}
\end{cases}
\end{equation}
\noindent and we will work only in these two bases for the remainder of this work.

We then see that from Equations (\ref{met:expandvol}) and (\ref{met:IA})-(\ref{met:tALS}) that if we want the total volume $\mathcal{V}$ to be large, then
\begin{equation}
\exists (i,j,k)\in I^{A}_{L}\times I^{A}\times I^{A}:\;\left(T^{A}\right)_{i}^{\;\; r}\left(T^{A}\right)_{j}^{\;\; s}\left(T^{A}\right)_{k}^{\;\; t}\kappa_{rst}\neq 0 \, . \label{met:vol}
\end{equation}
We also see from Equations (\ref{met:tauA}) and (\ref{met:IA})-(\ref{met:tauALS}) that
\begin{align}
&\exists(j,k)\in I^{A}_{L}\times I^{A}:\;\left(T^{B}\right)_{i}^{\;\; r}\left(T^{A}\right)_{j}^{\;\; s}\left(T^{A}\right)_{k}^{\;\; t}\kappa_{rst}\neq 0,\;\;\forall i\in I^{B}_{L}\label{met:LC}\\
&\left(T^{B}\right)_{i}^{\;\; r}\left(T^{A}\right)_{j}^{\;\; s}\left(T^{A}\right)_{k}^{\;\; t}\kappa_{rst}=0,\;\;\forall (i,j,k)\in I^{B}_{S}\times I^{A}_{L}\times I^{A} \, . \label{met:preSC}
\end{align}
\noindent But, if $\{J^{A}_{i}\}$ is a basis, then $\mathbf{T}^{A}$ must be full rank. This implies that we can write Equation~(\ref{met:preSC}) as
\begin{align}
\left(T^{B}\right)_{i}^{\;\; r}\left(T^{A}\right)_{j}^{\;\; s}\kappa_{rst}=0,\;\;\forall (i,j,t)\in I^{B}_{S}\times I^{A}_{L}\times I \, . \label{met:SC}
\end{align}

\subsection{K\"ahler Cone Condition}

\subsubsection{The Mori and K\"ahler Cones}

A K\"ahler manifold is defined as a symplectic manifold with a closed symplectic 2-form $J$, which is simultaneously consistent with an almost complex and Riemannian structure. The former imposes the constraint that $J$ is in fact a (1,1)-form, while the latter requires $J$ to be locally positive definite. This is directly related to the fact that the volume of an effective curve $\text{vol}(\mathcal{C})=\int_{\mathcal{C}}{J}>0$. By expanding $J=t^{i}J_{i}$ in a basis $\{J_{i}\}\in H^{1,1}(X)$, we ensure that it is indeed a closed (1,1)-form, however we must also constrain it to be positive definite. To make this explicit, we check that $J$ has positive intersection with every subvariety of complementary codimension, i.e. curves $\mathcal{C}$
\begin{equation}
\mathcal{K}(\mathcal{A})=\left\{J\in H^{1,1}(X)\left|\;\text{vol}(\mathcal{C})=\int\limits_{\mathcal{C}}{J}>0\right.\right\} \, . \label{Kcone}
\end{equation}
Thus, the allowed values of $J$ form a convex cone in the K\"ahler moduli space. The curves $\mathcal{C}$ then form a dual cone, known as the Mori cone $\mathcal{M}(\mathcal{A})\subset\text{Hom}\left(H^{1,1}(\mathcal{A}),\mathbb{Q}\right)\cong\mathbb{Q}^{h}$ which is generated\footnote{Again, for the sake of computational efficiency, we have relaxed the requirement of $J\in H^{1,1}(\mathcal{A})\cap H^{2}(\mathcal{A};\mathbb{Z})$ to $J\in H^{1,1}(\mathcal{A})\cap H^{2}(\mathcal{A};\mathbb{Q})$.} by a set of extremal rays $\mathcal{C}^{1},...,\mathcal{C}^{r}$ such that
\begin{equation}
\mathcal{M}(\mathcal{A})=\left\{\sum\limits_{i=1}^{r}{a_{i}\mathcal{C}^{i}}\mid a_{i}\in\mathbb{R}_{\geq 0}\right\} \, .
\end{equation}
These extremal rays can be regarded as linear functionals on the divisors, and can therefore easily be computed in terms of the toric divisors from symplectic moment polytope information provided in the Kreuzer-Skarke database. Then, given our original basis of divisor classes $\{J_{i}\}_{i\in I}$ of $H^{1,1}(X)\cong H^{1,1}(\mathcal{A})$ for favorable geometries, we can define the $r\times h$ K\"ahler cone matrix of intersection numbers between the generating curves $\mathcal{C}^{i}$ and the basis divisor classes
\begin{equation}
\mathbf{K}^{i}_{\;\; j}=\int\limits_{\mathcal{C}^{i}}{J_{j}} \, , \label{met:kahlermat}
\end{equation}
\noindent whose rows represent the generating curves, or equivalently, rays of the Mori cone $\mathcal{M}(\mathcal{A})$. Using this K\"ahler cone matrix, and referring to Equations~(\ref{met:basisexpansion}) and~(\ref{met:orig2basis}), we see that
\begin{equation}
\int\limits_{\mathcal{C}^{i}}{J}=t^{Aj}\int\limits_{\mathcal{C}^{i}}{J^{A}_{j}}=t^{Aj}\left(T^{A}\right)_{j}^{\;\; k}\int\limits_{\mathcal{C}^{i}}{J_{k}}=t^{Aj}\left(T^{A}\right)_{j}^{\;\; k}K^{i}_{\;\; k} \, ,
\end{equation}
\noindent where $J\in H^{1,1}(\mathcal{A})$ is the K\"ahler form on $\mathcal{A}$. If we want $J\in\mathcal{K}(\mathcal{A})$, then we must satisfy $\int\limits_{\mathcal{C}}{J}>0,\;\;\forall \mathcal{C}\in\mathcal{M}(\mathcal{A})$. This is equivalent to
\begin{equation}
0<\int\limits_{\mathcal{C}}{J}=\int\limits_{\sum\limits_{i}{a_{i}\mathcal{C}^{i}}}{J}=\sum\limits_{i}{a_{i}\int\limits_{\mathcal{C}^{i}}{J}},\text{  with  }a_{i}\in\mathbb{R}_{>0},\;\forall i\in I\, .
\end{equation}
Since this must be true for arbitrary $a_{i}$, then each term of the sum must satisfy the inequality independently
\begin{equation}
0<\int\limits_{\mathcal{C}^{i}}{J}=t^{Aj}\left(T^{A}\right)_{j}^{\;\; k}K^{i}_{\;\; k}=\left(K^{A}\right)^{i}_{\;\; j}t^{Aj},\;\;\forall i\in I \, . \label{met:kahler}
\end{equation}
\noindent This, then, is the set of conditions which must be satisfied in order for the K\"ahler form $J$ to lie within the K\"ahler cone. Unfortunately, this procedure only tells us the K\"ahler cone of the ambient toric variety $\mathcal{A}$, while that of the Calabi-Yau hypersurface may be larger. It is still, however, a sufficient condition.

In order to approximate better the full K\"ahler cone of the hypersurface, we have implemented the procedure, as described in our previous work \cite{Altman:2014bfa}, of gluing together the K\"ahler cones of all resolutions of $\mathcal{A}$ that are related by flops, and between which the hypersurface $X$ continues smoothly. It has been shown \cite{Szendroi2001} that in some cases, this procedure still results in a subcone of the full hypersurface K\"ahler cone. With some knowledge of the divisor structure, it can be further refined \cite{Cicoli:2012vw}, however, we will leave this to future work.

\subsubsection{Large and Small Cycle K\"ahler Cone Conditions}

Without loss of generality, we can always rearrange the rows of $\mathbf{K}^{A}=\mathbf{K}\left(\mathbf{T}^{A}\right)^{T}$ to put as many zero entries as possible in the lower left quadrant
\begin{equation}
\mathbf{K}^{A}=\begin{blockarray}{ccc|c}
\begin{block}{(ccc|c)}
 & & & \\
\mathbf{p}^{A}_{1} & \dots & \mathbf{p}^{A}_{N_{L}} & \ddots\\
 & & & \\\cline{1-4}
  & & & \\
 & \mathbf{0} & & \begin{array}{c}\left(\mathbf{q}^{A}_{1}\right)^{T}\\\vdots\\\left(\mathbf{q}^{A}_{m}\right)^{T}\end{array}\\
  & & & \\
\end{block}
\multicolumn{3}{c}{\upbracefill} & \multicolumn{1}{c}{\upbracefill} \\
\multicolumn{3}{c}{$\scriptstyle N_{L}$} & \multicolumn{1}{c}{$\scriptstyle h-N_{L}$} \\
\end{blockarray}
\begin{array}{c}\scalebrace{2}{r-m}\\\scalebrace{4}{m}\\\\\\\end{array},\hspace{1cm}0\leq m\leq r \, .
\end{equation}
\noindent Then, in the large volume limit where $\left(t^{A}\right)^{1},\dots,\left(t^{A}\right)^{N_{L}}\rightarrow\pm\infty$, the K\"ahler cone condition of Equation~(\ref{met:kahler}) becomes
\begin{align}
\sum\limits_{i=1}^{N_{L}}{\lim_{t^{Ai}\to\pm\infty}\mathbf{p}^{A}_{i}t^{Ai}}&>0\label{kahlerL1}\\
\notag\\
\mathbf{q}^{A}_{j}\cdot\left[\begin{array}{c}
\left(t^{A}\right)^{N_{L}+1}\\
\vdots\\
\left(t^{A}\right)^{h}
\end{array}\right]&>0,\;\;\forall j\in [1,m] \, . \label{kahlerS1}
\end{align}
For the first expression to be well-defined, each term must either be satisfied independently or be identically zero, so that
\begin{align}
&\lim_{t^{Ai}\to\pm\infty}\mathbf{p}^{A}_{i}t^{Ai}\geq 0,\;\;\forall i\in I^{A}_{L}\notag\\
\Rightarrow\hspace{0.5cm} &\mathbf{p}^{A}_{i}\geq\mathbf{0}\hspace{.5cm}\text{or}\hspace{.5cm}\mathbf{p}^{A}_{i}\leq\mathbf{0},\;\;\forall i\in I^{A}_{L}\notag\\
\Rightarrow\hspace{0.5cm} &\pm\mathbf{p}^{A}_{i}\geq\mathbf{0},\;\;\forall i\in I^{A}_{L} \, . \label{kahlerL2}
\end{align}

In order to satisfy Equation~(\ref{kahlerS1}), we first recognize that the rows of $\mathbf{K}^{A}$ are just the generating rays of the Mori cone $\mathcal{M}(\mathcal{A})\subset\mathbb{Q}^{h}$, as expressed in the $A$-basis of $H^{1,1}(X)$. Then, we see that $\text{cone}\left(\mathbf{q}^{A}_{1},\dots,\mathbf{q}^{A}_{m}\right)$ must be a convex subcone of at most dimension $h-N_{L}$. Then, defining the dual $\sigma^{\vee}$ to a $d$-dimensional convex cone $\sigma$ by
\begin{equation}
\sigma^{\vee}=\left\{\mathbf{n}\in\mathbb{Q}^{d}\left\vert\langle\mathbf{m},\mathbf{n}\rangle\geq 0,\;\;\forall\mathbf{m}\in\sigma\subset\mathbb{Q}^{d}\right.\right\}\, ,
\end{equation}
\noindent we see that the solution space of Equation~(\ref{kahlerS1}) is just the relative interior of the dual cone, where the inequality is strict
\begin{equation}
\left[\begin{array}{c}
\left(t^{A}\right)^{N_{L}+1}\\
\vdots\\
\left(t^{A}\right)^{h^{1,1}}
\end{array}\right]\in\text{relint}\left(\text{cone}\left(\mathbf{q}^{A}_{1},\dots,\mathbf{q}^{A}_{m}\right)^{\vee}\right) \, . \label{solkahlerS1}
\end{equation}
\noindent Thus, a solution exists if and only if
\begin{equation}
\text{dim}\left[\text{relint}\left(\text{cone}\left(\mathbf{q}^{A}_{1},\dots,\mathbf{q}^{A}_{m}\right)^{\vee}\right)\right]>0\, . \label{kahlerS2}
\end{equation}

\subsection{Homogeneity Condition}\label{sec:homogeneity}

The effective potential in the low-energy supergravity limit of a type IIB theory in the Large Volume Scenario (LVS) has exponential factors involving small cycle moduli that are proportional to $\mathcal{V}$, and can often be volatile unless the terms are carefully balanced. More specifically, to have a finite minimum, each term must be of the same order in $\mathcal{V}^{-1}$. We refer to this property as \textit{homogeneity} of the terms in the effective potential. This leads to a restrictive requirement on the K\"ahler potential, and in turn on the K\"ahler metric. Because the 4-cycle volumes obey the ordering $\tau^{B}_{i}\gg\tau^{B}_{j},\;\;\forall(i,j)\in I^{B}_{L}\times I^{B}_{S}$, all terms in the effective potential involving $\tau^{B}_{i}$ are exponentially suppressed for each $i\in I^{B}_{L}$. The requirement on the K\"ahler metric can then be expressed as

\begin{align}
\left(K^{-1}\right)_{ii}\sim \mathcal{V}h_{i}^{1/2}\left(\{\tau^{B}_{k}\}_{k\in I^{B}_{S}}\right),\;\;\forall i\in I^{B}_{S} \, ,
\end{align}
\noindent where the $\{h_{i}^{1/2}\}_{i\in I^{B}_{S}}$ are $h-N_{L}$ functions of degree-$1/2$ in the small 4-cycles $\{\tau^{B}_{k}\}_{k\in I^{B}_{S}}$. Now, we consider the expansion of the K\"ahler metric $\left(K^{-1}\right)_{ij}$ in $\mathcal{V}^{-1}$ (see Appendix~\ref{sec:appb} for details)~\cite{Bobkov:2004cy,Cicoli:2008va}
\begin{align}
\left(K^{-1}\right)_{ij}&=-4\mathcal{V}\left(\int\limits_{X}{J^{B}_{i}\wedge J^{B}_{j}\wedge J}\right)+4\tau^{B}_{i}\tau^{B}_{j}+\mathcal{O}\left(\mathcal{V}^{-1}\right)\notag\\
&=-4\mathcal{V}t^{Ak}\left(\int\limits_{X}{J^{B}_{i}\wedge J^{B}_{j}\wedge J^{A}_{k}}\right)+4\tau^{B}_{i}\tau^{B}_{j}+\mathcal{O}\left(\mathcal{V}^{-1}\right)\notag\\
&=-4\mathcal{V}t^{Ak}\left(T^{B}\right)_{i}^{\;\; r}\left(T^{B}\right)_{j}^{\;\; s}\left(T^{A}\right)_{k}^{\;\; t}\left(\int\limits_{X}{J_{r}\wedge J_{s}\wedge J_{t}}\right)+4\tau^{B}_{i}\tau^{B}_{j}+\mathcal{O}\left(\mathcal{V}^{-1}\right)\notag\\
&=-4\mathcal{V}t^{Ak}\left(T^{B}\right)_{i}^{\;\; r}\left(T^{B}\right)_{j}^{\;\; s}\left(T^{A}\right)_{k}^{\;\; t}\kappa_{rst}+4\tau^{B}_{i}\tau^{B}_{j}+\mathcal{O}\left(\mathcal{V}^{-1}\right)\, .
\end{align}
The diagonal elements of $\left(K^{-1}\right)_{ij}$ have the form
\begin{align}
\frac{\left(K^{-1}\right)_{ii}}{\mathcal{V}}&=-4t^{Aj}\left(T^{B}\right)_{i}^{\;\; r}\left(T^{B}\right)_{i}^{\;\; s}\left(T^{A}\right)_{j}^{\;\; t}\kappa_{rst}+4\frac{\left(\tau^{B}_{i}\right)^{2}}{\mathcal{V}}+\mathcal{O}\left(\mathcal{V}^{-1}\right)\, .
\end{align}
\noindent But, but by definition, $\tau^{B}_{i}\ll\mathcal{V},\;\;\forall i\in I^{B}_{S}$, so
\begin{align}
\frac{\left(K^{-1}\right)_{ii}}{\mathcal{V}}&\approx -4t^{Aj}\left(T^{B}\right)_{i}^{\;\; r}\left(T^{B}\right)_{i}^{\;\; s}\left(T^{A}\right)_{j}^{\;\; t}\kappa_{rst},\;\;\forall i\in I^{B}_{S}\, . \label{hfunctemp}
\end{align}
\noindent Then, because the 4-cycle volumes are quadratic in the 2-cycle volumes, we have found our degree-$1/2$ functions $\{h_{i}^{1/2}\}_{i\in I^{B}_{S}}$ from Equation~(\ref{hfunctemp})
\begin{align}
h_{i}^{1/2}\left(\{\tau^{B}_{j}\}_{j\in I^{B}_{S}}\right)=-4t^{Aj}\left(T^{B}\right)_{i}^{\;\; r}\left(T^{B}\right)_{i}^{\;\; s}\left(T^{A}\right)_{j}^{\;\; t}\kappa_{rst},\;\;\forall i\in I^{B}_{S}\, . \label{met:hfuncs}
\end{align}
\noindent By inspecting Equations~(\ref{met:tALS}) and~(\ref{met:hfuncs}), we find the following
\begin{align}
&\left(T^{B}\right)_{i}^{\;\; r}\left(T^{B}\right)_{i}^{\;\; s}\left(T^{A}\right)_{j}^{\;\; t}\kappa_{rst}=0,\;\;\forall (i,j)\in I^{B}_{S}\times I^{A}_{L}\label{met:redundSC}\\
\notag\\
&\exists j\in I^{A}_{S}:\;\left(T^{B}\right)_{i}^{\;\; r}\left(T^{B}\right)_{i}^{\;\; s}\left(T^{A}\right)_{j}^{\;\; t}\kappa_{rst}\neq 0,\;\;\forall i\in I^{B}_{S}\, . \label{met:hom}
\end{align}
\noindent Note that because $\kappa_{rst}$ is a symmetric tensor, Equation (\ref{met:redundSC}) is implied by Equation (\ref{met:SC}) and therefore redundant.

This homogeneity condition is critically important for finding a Swiss cheese solution with $N_{S}=1$. However, when $N_{S}>1$, the exponential factors in the effective potential have more degrees of freedom, and the necessity of this condition is loosened. However, it remains a sufficient condition in most circumstances, and we simply flag these cases in our scan when we encounter them, rather than constraining the search parameters.

\subsection{General List of Conditions}\label{sec:genconds}

In this section, we have compiled all the conditions necessary for $X$ to have a Swiss cheese solution in the large volume scenario. For ease of notation, we make the following definitions
\begin{align}
\kappa^{AAA}_{ijk}&=\left(T^{A}\right)_{i}^{\;\; r}\left(T^{A}\right)_{j}^{\;\; s}\left(T^{A}\right)_{k}^{\;\; t}\kappa_{rst}\notag\\
\kappa^{BAA}_{ijk}&=\left(T^{B}\right)_{i}^{\;\; r}\left(T^{A}\right)_{j}^{\;\; s}\left(T^{A}\right)_{k}^{\;\; t}\kappa_{rst}\notag\\
\kappa^{BA0}_{ijk}&=\left(T^{B}\right)_{i}^{\;\; r}\left(T^{A}\right)_{j}^{\;\; s}\kappa_{rsk}\notag\\
\kappa^{BBA}_{ijk}&=\left(T^{B}\right)_{i}^{\;\; r}\left(T^{B}\right)_{j}^{\;\; s}\left(T^{A}\right)_{j}^{\;\; t}\kappa_{rst} \, .
\end{align}
Then, in order for a Swiss cheese solution to exist with $N_{L}$ large 4-cycles, there must exist invertible\footnote{$\mathbf{T}^{A},\mathbf{T}^{B}\in GL_{h}(\mathbb{Q})$ implies that both are invertible: $\text{Det}\left(\mathbf{T}^{A}\right)\neq 0\;\;\text{ and }\;\;\text{Det}\left(\mathbf{T}^{B}\right)\neq 0$.} $A$- and $B$-bases such that

\begin{tabular}{l l l}
\\
1.&(Equation (\ref{met:vol}): Volume)&$\exists (i,j,k)\in I^{A}_{L}\times I^{A}\times I^{A}:\;\kappa^{AAA}_{ijk}\neq 0$\\
\\
2.&(Equation (\ref{met:LC}): Large Cycle)&$\exists(j,k)\in I^{A}_{L}\times I^{A}:\;\kappa^{BAA}_{ijk}\neq 0,\;\;\forall i\in I^{B}_{L}$\\
\\
3.&(Equation (\ref{met:SC}): Small Cycle)&$\kappa^{BA0}_{ijk}=0,\;\;\forall (i,j,k)\in I^{B}_{S}\times I^{A}_{L}\times I$\\
\\
4.&(Equation (\ref{met:hom}): Homogeneity)&$\exists j\in I^{A}_{S}:\;\kappa^{BBA}_{iij}\neq 0,\;\;\forall i\in I^{B}_{S}$\\
\\
5.&(Equation (\ref{kahlerL2}): K\"ahler Cone (L))&$\pm\mathbf{p}^{A}_{i}\geq\mathbf{0},\;\;\forall i\in I^{A}_{L}$\\
\\
6.&(Equation (\ref{kahlerS2}): K\"ahler Cone (S))&$\text{dim}\left[\text{relint}\left(\text{cone}\left(\mathbf{q}^{A}_{1},\dots,\mathbf{q}^{A}_{m}\right)^{\vee}\right)\right]>0$\\
\end{tabular}

\noindent where
\begin{equation*}
\mathbf{K}^{A}=\begin{blockarray}{ccc|c}
\begin{block}{(ccc|c)}
 & & & \\
\mathbf{p}^{A}_{1} & \dots & \mathbf{p}^{A}_{N_{L}} & \ddots\\
 & & & \\\cline{1-4}
  & & & \\
 & \mathbf{0} & & \begin{array}{c}\left(\mathbf{q}^{A}_{1}\right)^{T}\\\vdots\\\left(\mathbf{q}^{A}_{m}\right)^{T}\end{array}\\
  & & & \\
\end{block}
\multicolumn{3}{c}{\upbracefill} & \multicolumn{1}{c}{\upbracefill} \\
\multicolumn{3}{c}{$\scriptstyle N_{L}$} & \multicolumn{1}{c}{$\scriptstyle h-N_{L}$} \\
\end{blockarray}
\begin{array}{c}\scalebrace{2}{r-m}\\\scalebrace{4}{m}\\\\\\\end{array},\hspace{1cm}0\leq m\leq r \, ,
\end{equation*}
\noindent is the K\"ahler cone matrix after rotation into the $A$-basis.

\subsection{Special Case: Toric Swiss Cheese}\label{toriccase}

Each favorable toric Calabi-Yau threefold is endowed with a set of special 2-form toric classes $\{D_{i}\}_{i\in I^{\text{Toric}}}$ dual to the 4-cycle toric divisors, which descend directly from the ambient space $\mathcal{A}$. The K\"ahler moduli space $H^{1,1}(X)$ is always spanned by these toric 2-forms, up to some redundancy. We know, therefore, than any basis expansion of a point in the moduli space may be equivalently described as a linear combination of the toric 2-forms, though it will not be unique. Practically speaking, however, this form is advantageous for a scan since the ring structure of $H^{1,1}(X)$ has already been computed for these directly via toric methods and will not cost us anything. In addition, the redundancy of the toric divisors allows us to scan over multiple choices of basis (in particular, many will naturally be $\mathbb{Z}$-bases) simply by sampling subsets of the toric divisors. So, while there is no natural basis for our calculations, the toric divisor classes $\{D_{i}\}_{i\in I^{\text{Toric}}}$ form a natural ``pseudo-basis'' in spite of their redundancy.

A basis formed by a pure subset of the toric 2-forms will not always have a Swiss cheese solution. Even if such a solution exists, an arbitrary rotation may still be required. However, if we limit ourselves to the case in which some subset of the toric 2-forms is already a Swiss cheese basis (i.e. $\{J^{A}_{i}\}_{i\in I^{A}},\{J^{B}_{i}\}_{i\in I^{B}}\subset \{D_{i}\}_{i\in I^{\text{Toric}}}$), then our problem is reduced to a relatively simple combinatorical one. In order to see this, we define the injective maps
\begin{align}\label{eq:permute}
\alpha :\;&I^{A}\hookrightarrow I^{\text{Toric}}&\beta :\;&I^{B}\hookrightarrow I^{\text{Toric}}\notag\\
&J^{A}_{i}\mapsto D_{\alpha(i)}&&J^{B}_{i}\mapsto D_{\beta(i)} \, .
\end{align}
We also define the toric triple intersection tensor and the Mori cone matrix\footnote{The Mori cone matrix is essentially the same as the K\"ahler cone matrix, but expanded in the toric divisors rather than a basis. We give it this name because it is the object that is directly computed from torus invariant curves viewed as linear functionals relating the toric divisors.}
\begin{align}
d_{ijk}&=\int\limits_{X}{D_{i}\wedge D_{j}\wedge D_{k}}\\
\mathbf{M}^{i}_{\;\; j}&=\int\limits_{C^{i}}{D_{j}} \, .
\end{align}
Then, we can rewrite
\begin{align}
\kappa^{AAA}_{ijk}&=d_{\alpha(i)\alpha(j)\alpha(k)}\notag\\
\kappa^{BAA}_{ijk}&=d_{\beta(i)\alpha(j)\alpha(k)}\notag\\
\kappa^{BA0}_{ijk}&=d_{\beta(i)\alpha(j)k}\notag\\
\kappa^{BBA}_{ijk}&=d_{\beta(i)\beta(j)\alpha(k)}\notag\\
\left(\mathbf{K}^{A}\right)^{i}_{\;\; j}&=\mathbf{M}^{i}_{\;\; \alpha(j)} \, .
\end{align}
It is clear, then, that the conditions in Section~\ref{sec:genconds} become purely combinatoric in nature and take the form

\begin{tabular}{l l l}
\\
1.&(Equation (\ref{met:vol}): Volume)&$\exists (i,j,k)\in \alpha(I^{A}_{L})\times\alpha(I^{A})\times\alpha(I^{A}):\; d_{ijk}\neq 0$\\
\\
2.&(Equation (\ref{met:LC}): Large Cycle)&$\exists(j,k)\in \alpha(I^{A}_{L})\times\alpha(I^{A}):\; d_{ijk}\neq 0,\;\;\forall i\in \beta(I^{B}_{L})$\\
\\
3.&(Equation (\ref{met:SC}): Small Cycle)&$d_{ijk}=0,\;\;\forall (i,j,k)\in \beta(I^{B}_{S})\times\alpha(I^{A}_{L})\times I^{\text{Toric}}$\\
\\
4.&(Equation (\ref{met:hom}): Homogeneity)&$\exists j\in\alpha(I^{A}_{S}):\; d_{iij}\neq 0,\;\;\forall i\in\beta(I^{B}_{S})$\\
\\
5.&(Equation (\ref{kahlerL2}): K\"ahler Cone (L))&$\pm\mathbf{p}^{A}_{i}\geq\mathbf{0},\;\;\forall i\in\alpha(I^{A}_{L})$\\
\\
6.&(Equation (\ref{kahlerS2}): K\"ahler Cone (S))&$\text{dim}\left[\text{relint}\left(\text{cone}\left(\mathbf{q}^{A}_{1},\dots,\mathbf{q}^{A}_{m}\right)^{\vee}\right)\right]>0$\\
\end{tabular}

\noindent where
\begin{equation*}
\mathbf{M}^{A}=\begin{blockarray}{ccc|c}
\begin{block}{(ccc|c)}
 & & & \\
\mathbf{p}^{A}_{\alpha(1)} & \dots & \mathbf{p}^{A}_{\alpha(N_{L})} & \ddots\\
 & & & \\\cline{1-4}
  & & & \\
 & \mathbf{0} & & \begin{array}{c}\left(\mathbf{q}^{A}_{1}\right)^{T}\\\vdots\\\left(\mathbf{q}^{A}_{m}\right)^{T}\end{array}\\
  & & & \\
\end{block}
\multicolumn{3}{c}{\upbracefill} & \multicolumn{1}{c}{\upbracefill} \\
\multicolumn{3}{c}{$\scriptstyle N_{L}$} & \multicolumn{1}{c}{$\scriptstyle h-N_{L}$} \\
\end{blockarray}
\begin{array}{c}\scalebrace{2}{r-m}\\\scalebrace{4}{m}\\\\\\\end{array},\hspace{1cm}0\leq m\leq r \, .
\end{equation*}
Now, instead of solving a complex linear system for two arbitrary rotation matrices $\mathbf{T}^{A},\mathbf{T}^{B}\in GL_{h}(\mathbb{Q})$, we simply need to choose two subsets $\alpha(I^{A}),\beta(I^{B})\subset I^{\text{Toric}}$. Since the toric triple intersection tensor $d_{ijk}$ and the Mori cone matrix $\mathbf{M}^{i}_{\;\; j}$ are basis-independent, it is a simple combinatoric matter to search $d_{ijk}$ for subtensors that meet these constraints. If one is found, then we are done and the sets $\alpha(I^{A})$ and $\beta(I^{B})$ determine the bases\footnote{Again, even if the original basis $\{J_{i}\}_{i\in I}$ is a $\mathbb{Z}$-basis, it is not guaranteed in our analysis that $\{J^{A}_{i}\}_{i\in I^{A}}$ and $\{J^{B}_{i}\}_{i\in I^{B}}$ are as well.} $\{J^{A}_{i}\}_{i\in I^{A}}=\{D_{j}\}_{j\in\alpha(I^{A})}$ and $\{J^{B}_{i}\}_{i\in I^{B}}=\{D_{j}\}_{j\in \beta(I^{B})}$ for which there exists such a Swiss cheese solution. From toric methods, we can easily obtain the rectangular transformation matrix $\mathbf{R}$ given by
\begin{align}
D_{i}=R_{i}^{\;\; j}J_{j} \, .
\end{align}
\noindent Then, the rotation matrices $\mathbf{T}^{A}$ and $\mathbf{T}^{B}$ are simply defined by
\begin{align}
\left(T^{A}\right)_{i}^{\;\; j}=R_{\alpha(i)}^{\;\;\;\;\;\; j}\hspace{1cm}\text{and}\hspace{1cm}\left(T^{B}\right)_{i}^{\;\; j}=R_{\beta(i)}^{\;\;\;\;\;\; j} \, .
\end{align}

\section{Implementing Toric Swiss Cheese Detection}\label{sec:algorithm}

Given the combinatorial conditions set forward in the previous section, it is fairly straightforward to scan the database of toric Calabi-Yau threefolds\cite{Altmana} for Swiss cheese solutions. The procedure we use is as follows, but there are many variations.

\begin{enumerate}
\item
From the database, we can readily obtain the toric triple intersection tensor $d_{ijk}$, the Mori cone matrix $\mathbf{M}$, and the weight matrix $\mathbf{W}$. The latter is defined by the conditions $\sum\limits_{\rho=1}^{k}{\bm{W}_{r}^{\;\;\rho}\bm{n}_{\rho}}=\bm{0}$ and $\bm{W}\geq 0$, where the $k$ 4-dimensional vectors $\{\bm{n}_{1},...,\bm{n}_{k}\}$ are the vertices of the dual polytope.

\item
The Small Cycle condition reads
\begin{equation*}
d_{ijk}=0,\;\;\forall (i,j,k)\in \beta(I^{B}_{S})\times\alpha(I^{A}_{L})\times I^{\text{Toric}} \, .
\end{equation*}
This tells us that we can search for any row of any submatrix of $d_{ijk}$ that contains all zeroes, and the indices of those rows and submatrices give us all possible combinations of $\alpha(I^{A}_{L})$ and $\beta(I^{B}_{S})$.
\item
We then assemble all possible complementary sets of indices $\alpha(I^{A}_{S})$ and $\beta(I^{B}_{L})$ from among the $k$ toric divisor indices to get the full sets $\alpha(I^{A})$ and $\beta(I^{B})$, each of which contain $h$ total indices.
\item
We construct the submatrices $W_{\alpha(i)}^{\;\;\;\alpha(j)}$ and $W_{\beta(i)}^{\;\;\;\beta(j)}$ of the weight matrix and check that both are full rank, otherwise we have chosen redundant toric divisors.
\item
We then check the Volume condition
\begin{equation*}
\exists (i,j,k)\in \alpha(I^{A}_{L})\times\alpha(I^{A})\times\alpha(I^{A}):\; d_{ijk}\neq 0 \, .
\end{equation*}
\item
Given the Mori cone matrix $\mathbf{M}$ and the set of indices $\alpha(I^{A})$, we construct the submatrix $\mathbf{M}^{i}_{\;\;\alpha(j)}$ and reorder the rows until it takes the form
\begin{equation*}
\sbox0{$\begin{matrix}\mathbf{p}^{A}_{\alpha(1)}&\dots&\mathbf{p}^{A}_{\alpha(N_{L})}\end{matrix}$}
\sbox1{$\begin{matrix}\left(\mathbf{q}^{A}_{1}\right)^{T}\\\vdots\\\left(\mathbf{q}^{A}_{m}\right)^{T}\end{matrix}$}
\mathbf{M}^{i}_{\;\;\alpha(j)}=\left[\begin{array}{c|c}
\usebox{0}&\vphantom{\usebox{1}}\makebox[\wd1]{$\ddots$}\\
\hline
\\
\vphantom{\usebox{0}}\makebox[\ht1]{\large $\mathbf{0}$}&\usebox{1}
\end{array}\right],\hspace{1cm}0\leq m\leq h \, .
\end{equation*}
\item
Next, we check the Large Cycle K\"ahler cone condition
\begin{equation*}
\pm\mathbf{p}^{A}_{i}\geq\mathbf{0},\;\;\forall i\in\alpha(I^{A}_{L}) \, ,
\end{equation*}
\item
Then the Small Cycle K\"ahler cone condition
\begin{equation*}
\text{dim}\left[\text{relint}\left(\text{cone}\left(\mathbf{q}^{A}_{1},\dots,\mathbf{q}^{A}_{m}\right)^{\vee}\right)\right]>0 \, ,
\end{equation*}
\item
The Large Cycle condition
\begin{equation*}
\exists(j,k)\in \alpha(I^{A}_{L})\times\alpha(I^{A}):\; d_{ijk}\neq 0,\;\;\forall i\in \beta(I^{B}_{L}) \, ,
\end{equation*}
\item
And finally, if we choose to, we can also check the Homogeneity condition
\begin{equation*}
\exists j\in\alpha(I^{A}_{S}):\; d_{iij}\neq 0,\;\;\forall i\in\beta(I^{B}_{S})\, .
\end{equation*}
\item
If all the conditions in Section \ref{toriccase} are satisfied, then the sets of indices $\alpha(I^{A})$ and $\beta(I^{B})$ are converted into rotation matrices 
\begin{equation}\label{eq:permmat}
\left(T^{A}\right)_{i}^{\;\; j}=R_{\alpha(i)}^{\;\;\;\;\;\; j}\hspace{1cm}\text{and}\hspace{1cm}\left(T^{B}\right)_{i}^{\;\; j}=R_{\beta(i)}^{\;\;\;\;\;\; j}
\end{equation}
\noindent where
\begin{align}
D_{i}=R_{i}^{\;\; j}J_{j}\, .
\end{align}
\item
We also check whether the $A$- and $B$-bases are $\mathbb{Z}$-bases. This is the case if and only if the remaining redundant toric divisors all intersect each other smoothly at a point on the desingularized ambient toric variety $\mathcal{A}$, up to an action of the fundamental group.

\item
We repeat this procedure for $N_{L}=1,\dots,h-1$, so that at least one 4-cycle is always large and at least one 4-cycle is always small. The results are recorded in the database\cite{Altmana} as well.

\item
Finally, we can take multiple passes at the dataset, beginning with a randomly chosen $GL_{k}(\mathbb{Z})$ transformation on the toric divisors $\{D_{1},...,D_{k}\}$ each time. The full Swiss cheese solution set should begin to converge after many iterations, but it is unclear how slow that convergence should be. This is still a significant improvement over the method of solving the linear system for $\mathbf{T}^{A}$ and $\mathbf{T}^{B}$, as each loop will uncover a handful of solutions with purely combinatorial efficiency. We save this larger scan for a later work.
\end{enumerate}

\section{Swiss Cheese Classification}\label{sec:class}


In previous studies, the majority of Swiss cheese geometries have been constructed explicitly using a top down approach.
Here, working from a vast database of known candidate geometries~\cite{Altman:2014bfa,Altmana}, we attack the problem from the bottom up with the hope of identifying as many viable Swiss cheese vacua as possible.
Toward this end, in this section we lay out a scheme for categorizing Swiss cheese geometries with varying degrees of generality.

The K\"ahler moduli $t^{i}$ are the natural geometrical parameters on the Calabi-Yau threefold $X$, and it is a simple matter to write the volume form in terms of these as
\begin{equation}
\mathcal{V}=\frac{1}{3!}t^{i}t^{j}t^{k}\kappa_{ijk} \, , \label{eq:volplain}
\end{equation}
\noindent where the intersection tensor $\kappa_{ijk}$ encodes the Chow ring structure on $X$.
In the low energy ten dimensional supergravity limit, the relevant field parameters descending from $X$ are the complexified\footnote{The $b_{i}$ are axionic partners of the $\tau_{i}$ 4-cycle volumes.} 4-cycle volumes $T_{i}=\tau_{i}+ib_{i}$. There is a natural injective map from the 2-cycles to the 4-cycles via
\begin{equation}
t^{i}\mapsto\tau_{i}\equiv\frac{\partial\mathcal{V}}{\partial t^{i}} \, . \label{eq:map}
\end{equation}
Depending on the Chow ring structure hidden in $\mathcal{V}$, it may be possible to choose a basis in which the map is invertible, at least on some subset of $t^{i}$.
If so, then it is possible to write $\mathcal{V}$ explicitly in terms of the 4-cycle volumes (at least partially).
In this case, we say that $X$ is \textit{explicitly} Swiss cheese.

In addition, the Swiss cheese condition requires that some set of large 4-cycles determine the scale of the overall volume $\mathcal{V}$, while the remaining small 4-cycles determine the scale of the missing ``holes''.
This can be observed directly from the form of $\mathcal{V}$ when each 4-cycle volume $\tau_{i}$ contributes independently as its own term, such that
\begin{equation}
\mathcal{V}=\sum\limits_{i=1}^{h^{1,1}}{\lambda_{i}\tau_{i}^{3/2}} \, . \label{eq:diag}
\end{equation}
\noindent When this is the case, we say that the volume is \textit{diagonalized}, as there are no mixed terms.
The conditions set forward in Section~\ref{sec:genconds} guarantee that $X$ obeys the Swiss cheese condition, but even when $X$ is explicit, it is not always possible to find a basis that makes Equation~(\ref{eq:diag}) manifest.
When it is possible, though, we say that $X$ is \textit{diagonal}.

Finally, a Swiss cheese geometry $X$ that is both maximally explicit and maximally diagonal has special properties, and we refer to it as a \textit{strong} Swiss cheese geometry.
In any other case, $X$ is said to be \textit{weak}.

\begin{table}[!t]
\centering
\begin{tabular}{|r||r|c|c||c|}
\hline
 & & \multicolumn{2}{c||}{\textbf{Weak}} & \\
\hline
\hline
 & & \textbf{Implicit} & \textbf{Partially Explicit} & \textbf{Explicit} \\
\hline
\multirow{2}{*}{\textbf{Weak}} & \textbf{Non-Diagonal} & $\left[f^{3}\right]$ & $f^{3},\left[f^{2}g^{1/2},f^{2}\tau^{1/2},fg,f\tau,g^{3/2}\right]$ & $\left[g^{3/2}\right]$ \\
\hhline{~----}
 & \textbf{Partially Diagonal} & -- & $f^{3},f^{2}g^{1/2},f^{2}\tau^{1/2},fg,f\tau ,g^{3/2},\left[\tau^{3/2}\right]$ & $g^{3/2},\left[\tau^{3/2}\right]$ \\
\hline
\hline
 & \textbf{Diagonal} & -- & -- & \begin{minipage}{1in}\centering$\left[\tau^{3/2}\right]$ (\textbf{Strong})\end{minipage} \\
\hline
\end{tabular}
\caption{Classification of the allowed forms of monomial terms $M_{i}$ in the volume polynomial $\mathcal{V}=\sum\limits_{i}{\lambda_{i}M_{i}}$, with at least one $M_{i}$ corresponding to a term in square brackets.}
\label{table:class1}
\end{table}

In order to give a more thorough classification, we first define the monomial functions

\begin{itemize}
\item $f^{d}\equiv f^{d}(t^{1},\dots,t^{h})$: a degree $d$ monomial in the 2-cycle volumes.
\item $g^{d}\equiv g^{d}(\tau_{1},\dots,\tau_{h})$: a degree $d$ monomial in the 4-cycle volumes.
\end{itemize}

\noindent With this notation, Table~\ref{table:class1} enumerates the monomial forms $M_{i}$ that can appear in the expression for the overall volume
\begin{equation}
\mathcal{V}=\sum\limits_{i}{\lambda_{i}M_{i}}\, . \label{eq:monom}
\end{equation}
\noindent Note that a special case of $g^d$ occurs when the case in question is a function of only one 4-cycle volume.
In these cases we have replaced $g^d$ with $\tau^d$ in Table~\ref{table:class1}.

The volume is naturally expressed in terms of 2-cycles as a sum of monomials of the form $f^3$, as in Equation~(\ref{eq:volplain}).
If none of the maps in Equation~(\ref{eq:map}) are invertible, then this form of the volume must remain \textit{implicit} only.
If one or more of the maps can be inverted, then mixed terms involving momomials from the various terms $g^d$ are possible.
Thus, a partially explicit case \textit{may} contain terms that remain in the form $f^3$, but \textit{must} contain terms involving $g^d$.
A completely explicit case involves a sum of monomials from the set $g^{3/2}$ \textit{only}.
We note that a fully explicit form for the volume may not always display the geometrical properties of the Calabi-Yau manifold most clearly.
For example, a K$3$-fibration is often evidenced by terms of the form $f\tau$ in the volume~\cite{Cicoli:2011it}.



We now focus our attention on the lower-right quadrant of Table \ref{table:class1}.
These are the cases that are designated as partially, or fully, diagonal.
When restricted to these cases, we can write the most general volume form in a new basis $C$ as
\begin{align}
\mathcal{V}=f^{3}+f^{2}g^{1/2}\left(\tau^{C}_{1},...,\tau^{C}_{h}\right)+fg\left(\tau^{C}_{1},...,\tau^{C}_{h}\right)+g^{3/2}\left(\tau^{C}_{1},...,\tau^{C}_{h}\right)+\sum\limits_{i=1}^{n}{\lambda^{C}_{i}\left(\tau^{C}_{i}\right)^{3/2}} \, , \label{eq:partdiag}
\end{align}
\noindent with $n\leq h$. Thus, a partially diagonal volume form \textit{may} contain factors from the general terms $f^d$ and $g^d$, but \textit{must} contain at least one term of the form $\tau^{3/2}$.
A circumstance such as Equation~(\ref{eq:diag}) is both fully explicit (no terms involving 2-cycle volumes) and fully diagonal. 

In contrast, consider the general form of the volume in terms of the K\"ahler moduli given by Equation (\ref{met:expandvol}):
\begin{align}
\mathcal{V}&=\frac{1}{3!}t^{Ci}t^{Cj}t^{Ck}\left(T^{C}\right)_{i}^{\;\; r}\left(T^{C}\right)_{j}^{\;\; s}\left(T^{C}\right)_{k}^{\;\; t}\kappa_{rst}\notag\\
&=\frac{1}{3!}t^{Ci}t^{Cj}t^{Ck}\kappa^{CCC}_{ijk} \, ,
\label{voltC}
\end{align}
\noindent where $\kappa^{CCC}_{ijk}=\left(T^{C}\right)_{i}^{\;\; r}\left(T^{C}\right)_{j}^{\;\; s}\left(T^{C}\right)_{k}^{\;\; t}\kappa_{rst}$. Then, we can recover the 4-cycle volumes
\begin{align}
\tau^{C}_{i}&=\frac{\partial\mathcal{V}}{\partial t^{Ci}}=\frac{1}{2}t^{Cj}t^{Ck}\kappa^{CCC}_{ijk}
\label{tauC}
\end{align}
\noindent and rewrite the volume as
\begin{align}
\mathcal{V}&=\frac{1}{3}t^{Ci}\tau^{C}_{i} .
\label{volttauC}
\end{align}
\noindent We can then scan the database of Calabi-Yau vacua for cases in which the Chow ring structure allows for the identification
\begin{align}
\tau^{D}_{i}=\left(T^{D}\right)_{i}^{\;\; j}\tau_{j}=\left(T^{D}\left(T^{C}\right)^{-1}\right)_{i}^{\;\;j}\tau^{C}_{j}=\frac{1}{9(\lambda_{i}^{C})^{2}}t^{Ci}t^{Ci},\;\;\text{ $\forall i\leq n$} \, ,
\label{tauDfromtC}
\end{align}
\noindent When this is the case, the volume takes the explicit form
\begin{align}
\mathcal{V}&=\sum\limits_{i=1}^{h}{\pm\lambda_{i}^{C}\tau^{C}_{i}\sqrt{\tau^{D}_{i}}}
\label{volttauCD}
\end{align}
\noindent where the sign of each coefficient $\lambda_{i}^{C}$ can be fixed by K\"ahler cone and non-negative volume considerations. Furthermore, when the $C$- and $D$-bases coincide, the volume can be written in the \textit{diagonal} form
\begin{align}
\mathcal{V}&=\sum\limits_{i=1}^{h}{\pm\lambda_{i}^{C}\left(\tau^{C}_{i}\right)^{3/2}} .
\label{volttauCDdiag}
\end{align}
In this case, the LVS vacuum takes the form of a ``strong'' Swiss cheese compactification, in which terms with negative sign punch out ``holes'' in an overall volume. Comparing Equations~(\ref{tauC}) and (\ref{tauDfromtC}), we see that
\begin{equation}
\kappa^{CCC}_{ijk}=\left\{\begin{array}{cl}
\frac{2}{9\left(\lambda^{C}_{i}\right)^{2}} \, , & i=j=k\leq n \\
\\
0 \, , & i,j,k\leq n \\
\\
\text{Undetermined}\; ,&\text{otherwise}
\end{array}\right.
\label{kC}
\end{equation}
\noindent Therefore, we see that in the $C$-basis, $\kappa^{CCC}_{ijk}=\left(T^{C}\right)_{i}^{r}\left(T^{C}\right)_{j}^{s}\left(T^{C}\right)_{k}^{t}\kappa_{rst}$ is a partially-diagonal, rank three tensor.
In fact, if $n=h$, then $\kappa^{CCC}_{ijk}$ is fully diagonal, and $X$ is a strong Swiss cheese geometry.
This result is derived via similar methods in~\cite{Cicoli:2011it}.

It must be noted carefully, however, that the above procedure is not exhaustive in identifying explicit Swiss cheese cases. It is clear from Equation (\ref{tauDfromtC}) that in this case, the maps of Equation (\ref{eq:map}) can be inverted with the specific form
\begin{equation}
\tau^{C}_{i}\mapsto t^{Ci}=\sqrt{\sum\limits_{j=1}^{h}{a_{i}^{\;\; j}\tau^{C}_{j}}} \, .
\end{equation}
\noindent This is, in fact, a very restrictive condition and will fail to detect many potentially interesting solutions. In Section \ref{sec:examples}, we showcase a specific example that is both an explicit and a toric Swiss cheese manifold. While the toric methods of Section \ref{sec:algorithm} allowed us to identify it as a solution, the methods of this section were insufficient to detect it as an explicit case. As a result, a more robust algorithm for inverting the maps of Equation (\ref{eq:map}) is currently in development.

\section{Example Moduli Stabilization: Toric Swiss Cheese with $N_{L}=N_{S}=2$}\label{sec:examples}
\label{sec:minim}


We choose an example from our database at \url{www.rossealtman.com} with $h^{1,1}(X)=4,\; h^{2,1}(X)=94,\; \chi(X)=-180$ and database indexes
\begin{equation*}
\begin{array}{|c|c|}
\hline
\text{Polytope ID}&\text{Geometry ID}\\
\hline
1145&1\\
\hline
\end{array}
\end{equation*}
The intersection numbers and K\"ahler cone matrix in the original bases are given by
\begin{align}
I_{3}=&J_1^2 J_2-3 J_1 J_2^2-9 J_2^3+2 J_1^2 J_3+6 J_1 J_2 J_3+6 J_1 J_3^2+18 J_2 J_3^2\notag\\
&+18 J_3^3+J_1^2 J_4-3 J_1 J_4^2+9 J_4^3 \, , \\
\notag\\
\mathbf{K}=&\left(\begin{array}{cccc}
 0 & 1 & 0 & 1 \\
 1 & 0 & 0 & -3 \\
 0 & 0 & 0 & -1 \\
 0 & -1 & 1 & 0 \\
\end{array}\right) \, .
\end{align}
This Calabi-Yau geometry was found, as a result of our toric Swiss cheese scan presented in Section~\ref{sec:algorithm}, to have a Swiss cheese solution with $N_{L}=N_{S}=2$, with original basis, $A$-basis, and $B$-basis given by
\begin{align}
J_{1}=D_{3}&,& J_{2}=D_{6}&,&J_{3}=D_{7}&,&J_{4}=D_{8} \, ,\\
\notag\\
J^{A}_{1}=D_{5}&,& J^{A}_{2}=D_{7}&,& J^{A}_{3}=D_{1}&,& J^{A}_{4}=D_{4} \, ,\\
\notag\\
J^{B}_{1}=D_{1}&,& J^{B}_{2}=D_{5}&,& J^{B}_{3}=D_{4}&,& J^{B}_{4}=D_{8} \, .
\end{align}
The toric divisors have independent Hodge numbers $h^{\bullet}=\{h^{0,0},h^{0,1},h^{0,2},h^{1,1}\}$ given by\footnote{In order to determine the Hodge number of an individual divisor on the Calabi-Yau threefold, we use the Koszul extension to the \texttt{cohomCalg} package \cite{Blumenhagen:2010pv, cohomCalg:Implementation} with the \texttt{HodgeDiamond} module.}
\begin{align}
h^{\bullet}(D_{1})=h^{\bullet}(D_{2})=h^{\bullet}(D_{3})&=\{1,0,2,30\}\notag\\
h^{\bullet}(D_{4})=h^{\bullet}(D_{8})&=\{1,0,0,1\}\notag\\
h^{\bullet}(D_{5})&=\{1,0,1,20\}\\
h^{\bullet}(D_{6})&=\{1,0,0,19\}\notag\\
h^{\bullet}(D_{7})&=\{1,0,10,92\}\notag \, .
\end{align}
Since the $B$-basis separates 4-cycles into large and small volumes given by $\tau^{B}_{i}=\frac{1}{2!}\int_{J^{B}_{i}}{J\wedge J}$, this tells us immediately that the two small volume divisors $J^{B}_{3}$ and $J^{B}_{4}$ are both dP$_{0}$ blowup cycles, while $J^{B}_{2}$ is a $K3$ fiber. Then, $J^{B}_{3}$ and $J^{B}_{4}$ are precisely the divisors desired to host the non-perturbative contributions to the superpotential (due to E3-instantons or gaugino condensation on a stack of D7 branes) required to stabilize some of the K\"ahler moduli using the LVS prescription.

\begin{figure}[!t]
\centering
\includegraphics[width=\textwidth]{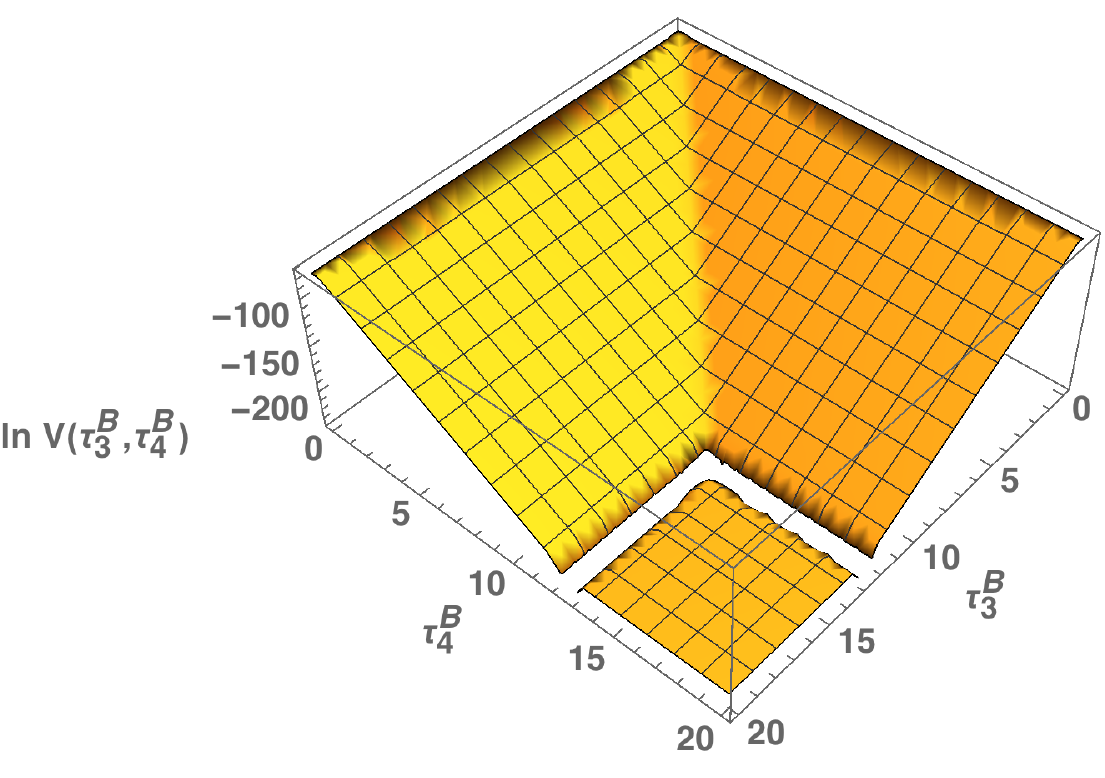}
\caption{Plot of $\text{ln }V\left(\tau^{B}_{3},\tau^{B}_{4}\right)$ with $\langle\mathcal{V}\rangle\sim 10^{32}$ and reasonable values for $a_{3}=a_{4}=2\pi$, $A_{1}=A_{2}=1$, $W_{GVW}=1$, and $\xi =-\frac{\chi (X)\zeta (3)}{2}$. }
\label{fig:t3t4}
\end{figure}

\begin{figure}[!h]
\centering
\includegraphics[width=\textwidth]{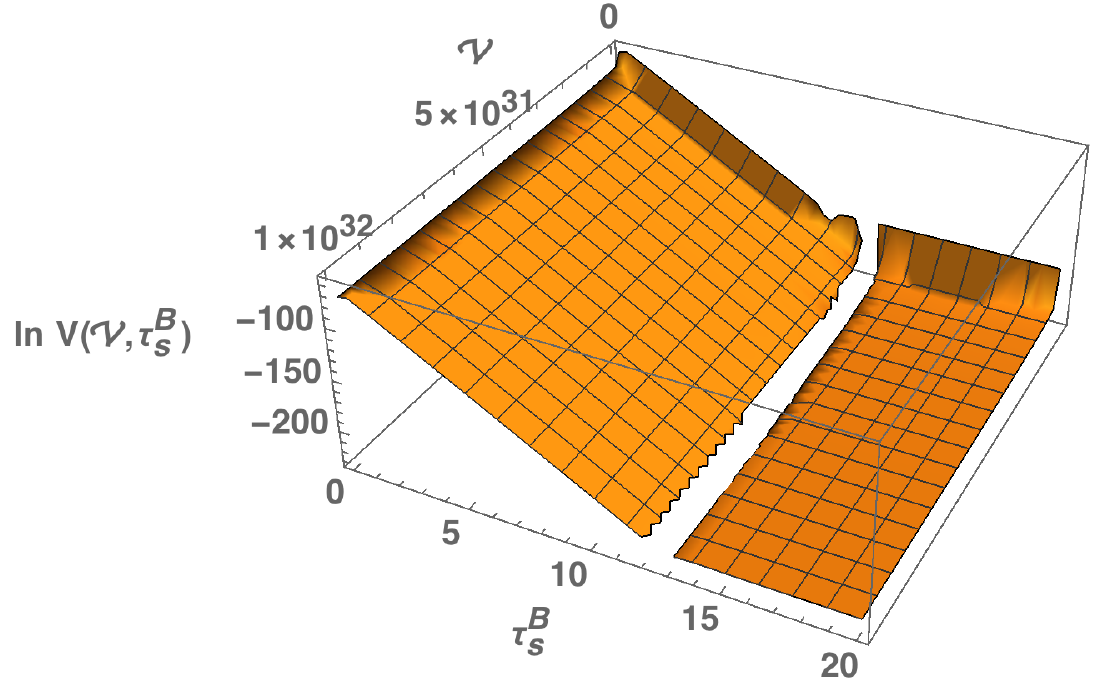}
\caption{Plot of $\text{ln }V\left(\mathcal{V},\tau^{B}_{s}\right)$ with $\tau^{B}_{s}:=\tau^{B}_{3}=\tau^{B}_{4}$, and reasonable values for $a_{3}=a_{4}=2\pi$, $A_{1}=A_{2}=1$, $W_{GVW}=1$, and $\xi =-\frac{\chi (X)\zeta (3)}{2}$. }
\label{fig:vts}
\end{figure}

Using Equation (\ref{eq:permmat}) and the relations between toric divisors, we find the rotation matrices
\begin{align}
\mathbf{T}^{A}=\left(\begin{array}{cccc}
-3 & 1 & 1 & -1\\
0 & 0 & 1 & 0\\
1 & 0 & 0 & 0\\
-3 & 0 & 1 & -1
\end{array}\right)\hspace{5mm}\text{and}\hspace{5mm}\mathbf{T}^{B}=\left(\begin{array}{cccc}
1 & 0 & 0 & 0\\
-3 & 1 & 1 & -1\\
-3 & 0 & 1 & -1\\
0 & 0 & 0 & 1
\end{array}\right) \, .
\end{align}
We can use these to rotate the intersection tensor into the $AAA$, $BAA$, and $BBA$ configurations
\begin{align}
\kappa^{AAA}_{ijk}&=\left(T^{A}\right)_{i}^{\;\; r}\left(T^{A}\right)_{j}^{\;\; s}\left(T^{A}\right)_{j}^{\;\; t}\kappa_{rst}\notag\\
\kappa^{BAA}_{ijk}&=\left(T^{B}\right)_{i}^{\;\; r}\left(T^{A}\right)_{j}^{\;\; s}\left(T^{A}\right)_{j}^{\;\; t}\kappa_{rst}\notag\\
\kappa^{BBA}_{ijk}&=\left(T^{B}\right)_{i}^{\;\; r}\left(T^{B}\right)_{j}^{\;\; s}\left(T^{A}\right)_{j}^{\;\; t}\kappa_{rst} \, .
\end{align}
The intersection numbers in this configuration are given by
\begin{align}
I^{AAA}_{3}=&18 J_1{}^A \left(J_2{}^A\right){}^2+18 \left(J_2{}^A\right){}^3+6 J_1{}^A J_2{}^A J_3{}^A+6 \left(J_2{}^A\right){}^2 J_3{}^A+2 J_1{}^A \left(J_3{}^A\right){}^2\notag\\
&+2 J_2{}^A \left(J_3{}^A\right){}^2+\left(J_3{}^A\right){}^2 J_4{}^A-3 J_3{}^A \left(J_4{}^A\right){}^2+9 \left(J_4{}^A\right){}^3\, , \\
\notag\\
I^{BAA}_{3}=&6 J_1{}^A J_1{}^B J_2{}^A+6 J_1{}^B \left(J_2{}^A\right){}^2+18 \left(J_2{}^A\right){}^2 J_2{}^B+2 J_1{}^A J_1{}^B J_3{}^A+2 J_1{}^B J_2{}^A J_3{}^A\notag\\
&+6 J_2{}^A J_2{}^B J_3{}^A+2 J_2{}^B \left(J_3{}^A\right){}^2+\left(J_3{}^A\right){}^2 J_3{}^B+J_1{}^B J_3{}^A J_4{}^A-3 J_3{}^A J_3{}^B J_4{}^A\notag\\
&-3 J_1{}^B \left(J_4{}^A\right){}^2+9 J_3{}^B \left(J_4{}^A\right){}^2+\left(J_3{}^A\right){}^2 J_4{}^B\, ,\\
\notag\\
I^{BBA}_{3}=&2 J_1{}^A \left(J_1{}^B\right){}^2+2 \left(J_1{}^B\right){}^2 J_2{}^A+6 J_1{}^B J_2{}^A J_2{}^B+2 J_1{}^B J_2{}^B J_3{}^A+J_1{}^B J_3{}^A J_3{}^B\notag\\
&-3 J_3{}^A \left(J_3{}^B\right){}^2+\left(J_1{}^B\right){}^2 J_4{}^A-3 J_1{}^B J_3{}^B J_4{}^A+9 \left(J_3{}^B\right){}^2 J_4{}^A+J_1{}^B J_3{}^A J_4{}^B\notag\\
&-3 J_3{}^A \left(J_4{}^B\right){}^2\, .
\end{align}
We can then write out the $\tau^{B}_{i}$ in terms of the $t^{Ai}$ using
\begin{align}
\tau^{B}_{i}=\frac{1}{2!}t^{Aj}t^{Ak}\kappa^{BAA}_{ijk}
\end{align}
\noindent and we get
\begin{align}\label{eq:BtoA}
\tau^{B}_{1}&=3\left(t^{A2}\right)^{2}+2t^{A2}t^{A3}+2t^{A1}\left(3t^{A2}+t^{A3}\right)+t^{A3}t^{A4}-\frac{3}{2}\left(t^{A4}\right)^{2}\notag\\
\tau^{B}_{2}&=\left(3t^{A2}+t^{A3}\right)^{2}\notag\\
\tau^{B}_{3}&=\frac{1}{2}\left(t^{A3}-3t^{A4}\right)^{2}\\
\tau^{B}_{4}&=\frac{1}{2}\left(t^{A3}\right)^{2} \, . \notag
\end{align}
Thus, we see that the $B$-basis is at least partially explicit. We can invert the perfect squares to get
\begin{align}\label{eq:invert}
3t^{A2}+t^{A3}&=\pm\sqrt{\tau^{B}_{2}}\notag\\
\frac{1}{\sqrt{2}}\left(t^{A3}-3t^{A4}\right)&=\pm\sqrt{\tau^{B}_{3}}\\
\frac{t^{A3}}{\sqrt{2}}&=\pm\sqrt{\tau^{B}_{4}}\, . \notag
\end{align}
We can fix the signs on the right hand side by computing the K\"ahler cone in the $A$-basis
\begin{align}
\mathbf{K}^{A}=\mathbf{K}\left(\mathbf{T}^{A}\right)^{T}=\left(
\begin{array}{cccc}
 0 & 0 & 0 & -1 \\
 0 & 0 & 1 & 0 \\
 1 & 0 & 0 & 1 \\
 0 & 1 & 0 & 1 \\
\end{array}
\right)\, ,
\end{align}
\noindent with $\left(K^{A}\right)_{i}^{\;\; j}t^{Aj}>0$, so that
\begin{align}
t^{A4}<0,\;\;\;\;\; t^{A3}>0,\;\;\;\;\; t^{A1}+t^{A4}>0,\;\;\;\;\; t^{A2}+t^{A4}>0\, .
\end{align}
This fixes the signs in Equation (\ref{eq:invert}) to be $(+,+,+)$. Solving the rest of Equation (\ref{eq:BtoA}), we get the rather messy result
\begin{align}\label{eq:invertfull}
t^{A}_{1}&=\frac{1}{6\sqrt{\tau^{B}_{2}}}\left(3\tau^{B}_{1}-\tau^{B}_{2}+\tau^{B}_{3}+\tau^{B}_{4}\right)\notag\\
t^{A}_{2}&=\frac{1}{3}\left(\sqrt{\tau^{B}_{2}}-\sqrt{2\tau^{B}_{4}}\right)\notag\\
t^{A}_{3}&=\sqrt{2\tau^{B}_{4}}\\
t^{A}_{4}&=\frac{\sqrt{2}}{3}\left(\sqrt{\tau^{B}_{4}}-\sqrt{\tau^{B}_{3}}\right)\, . \notag
\end{align}
Substituting these into the expression for volume, we get
\begin{align}\label{eq:explicitvol}
\mathcal{V}&=\frac{1}{3!}t^{Ai}t^{Aj}t^{Ak}\kappa^{AAA}_{ijk}\notag\\
&=\frac{1}{18}\left[9\tau^{B}_{1}\sqrt{\tau^{B}_{2}}+3\sqrt{\tau^{B}_{2}}\left(\tau^{B}_{3}+\tau^{B}_{4}\right)-\left(\tau^{B}_{2}\right)^{3/2} -2\sqrt{2}\left(\left(\tau^{B}_{3}\right)^{3/2}+\left(\tau^{B}_{4}\right)^{3/2}\right)\right]\, .
\end{align}
Thus, we have determined that this is an explicit and partially diagonal Swiss cheese solution. In order to stabilize the K\"ahler moduli, we must write down the effective potential and find a stable AdS minimum, which can later be uplifted. To find the form of the potential, we need to know the inverse K\"ahler metric. This is, to leading order in $\mathcal{V}^{-1}$ (see\footnote{Note that in Appendix \ref{sec:appb}, the inverse K\"ahler metric for the K\"ahler moduli is denoted $\left(\tilde{\cK}_{T}^{-1}\right)_{i\bar{j}}$, while here we refer to it simply at $\left(\cK^{-1}\right)_{i\bar{j}}$.} Equations \ref{eq:AVinv} and \ref{eq:BKinvT}), given by
\begin{align}
\left(\mathcal{K}^{-1}\right)_{i\bar{j}}&=-4\mathcal{V}\kappa^{BBA}_{ijk}t^{Ak}\notag\\
&=4\sqrt{2}\mathcal{V}\left(
\begin{array}{cccc}
 \frac{\left(\sqrt{\tau^{B}_{3}}+\sqrt{\tau^{B}_{4}}\right)}{3} -\frac{\left(3 \tau^{B}_{1}+\tau^{B}_{2}+\tau^{B}_{3}+\tau^{B}_{4}\right)}{3 \sqrt{2\tau^{B}_{2}}} & -\sqrt{2\tau^{B}_{2}} & -\sqrt{\tau^{B}_{3}} & -\sqrt{\tau^{B}_{4}} \\
 -\sqrt{2\tau^{B}_{2}} & 0 & 0 & 0 \\
 -\sqrt{\tau^{B}_{3}} & 0 & 3 \sqrt{\tau^{B}_{3}} & 0 \\
 -\sqrt{\tau^{B}_{4}} & 0 & 0 & 3 \sqrt{\tau^{B}_{4}} \\
\end{array}
\right)\, .
\end{align}
From this form of the inverse K\"ahler metric, the effective potential takes the form
\begin{align}
V\left(\mathcal{V},\tau^{B}_{3},\tau^{B}_{4}\right)&=\frac{a_{3}^2 |A_{3}|^2 \left(\mathcal{K}^{-1}\right)_{33} e^{-2 a_{3} \tau^{B}_{3}}}{2\mathcal{V}^{2}}+\frac{a_{4}^2 |A_{4}|^2 \left(\mathcal{K}^{-1}\right)_{44} e^{-2 a_{4} \tau^{B}_{4}}}{2\mathcal{V}^{2}}\notag\\
&\hspace{7mm}+\frac{2 a_{3}a_{4} |A_{3}A_{4}| \left(\mathcal{K}^{-1}\right)_{34} e^{-\left(a_{3} \tau^{B}_{3}+a_{4} \tau^{B}_{4}\right)}}{2\mathcal{V}^{2}}-\frac{2 a_{3} |A_{3}W_{GVW}| \tau^{B}_{3} e^{-a_{3} \tau^{B}_{3}}}{\mathcal{V}^2}\notag\\
&\hspace{7mm}-\frac{2 a_{4} |A_{4}W_{GVW}| \tau^{B}_{4} e^{-a_{4} \tau^{B}_{4}}}{\mathcal{V}^2}+\frac{3 \xi  |W_{GVW}|^2}{8 \mathcal{V}^3}\notag\\
&=\frac{6\sqrt{2}a_{3}^2 |A_{3}|^2 \sqrt{\tau^{B}_{3}} e^{-2 a_{3} \tau^{B}_{3}}}{\mathcal{V}}+\frac{6\sqrt{2} a_{4}^2 |A_{4}|^2 \sqrt{\tau^{B}_{4}} e^{-2 a_{4} \tau^{B}_{4}}}{\mathcal{V}}-\frac{2 a_{3} |A_{3}W_{GVW}| \tau^{B}_{3} e^{-a_{3} \tau^{B}_{3}}}{\mathcal{V}^2}\notag\\
&\hspace{7mm}-\frac{2 a_{4} |A_{4}W_{GVW}| \tau^{B}_{4} e^{-a_{4} \tau^{B}_{4}}}{\mathcal{V}^2}+\frac{3 \xi  |W_{GVW}|^2}{8 \mathcal{V}^3}\, .
\end{align}
We attempt to plot $\text{ln }V\left(\tau^{B}_{3},\tau^{B}_{4}\right)$, using an estimate of $\langle\mathcal{V}\rangle\sim 10^{32}$, and reasonable values for $a_{3}=a_{4}=2\pi$, $A_{1}=A_{2}=1$, $W_{GVW}=1$, and $\xi =-\frac{\chi (X)\zeta (3)}{2}$, in Figure~\ref{fig:t3t4}.
Where the logarithm gets cut off, the potential has gone negative. This gives us an AdS minimum. We notice that the potential is symmetric in $\tau^{B}_{3}$ and $\tau^{B}_{4}$, so we can choose the direction where they are equal, i.e. $\tau^{B}_{s}:\tau^{B}_{3}=\tau^{B}_{4}$. Then we can plot the potential in terms of $\tau^{B}_{s}$ and $\mathcal{V}$, as in Figure~\ref{fig:vts}.
Again, we see the AdS minimum. With $\tau^{B}_{3}$ and $\tau^{B}_{4}$ identified, the potential takes the form
\begin{align}
V\left(\mathcal{V},\tau_{s}\right)&=\frac{12 \sqrt{2} a_{s}^2 |A_{s}|^2 \sqrt{\tau^{B}_{s}} e^{-2 a_{s} \tau^{B}_{s}}}{\mathcal{V}}-\frac{4 a_{s} |A_{s}W_{GVW}| \tau^{B}_{s} e^{-a_{s} \tau^{B}_{s}}}{\mathcal{V}^2}+\frac{3 \xi  |W_{GVW}|^2}{8 \mathcal{V}^3}\, .
\end{align}
In fact, this is exactly the form of Equation (\ref{eq:standpot}), the potential for $N_{S}=1$ with $A_{s}\mapsto 2A_{s}$ and $c=\frac{3}{\sqrt{2}}$. Using Equations (\ref{eq:taumin}) and (\ref{eq:volmin}) to find the minima, we arrive at
\begin{align}
\left\langle\tau^{B}_{s}\right\rangle&\simeq\frac{1}{4}\left(\frac{3c\chi (X)\zeta (3)}{4}\right)^{2/3}\simeq 12.3 \, ,\\
\notag\\
\left\langle\cV\right\rangle&\simeq\frac{\left\lvert W_{\text{GVW}}\right\rvert}{2ca_{s}\left\lvert A_{s}\right\rvert}\sqrt{\tau^{B}_{s}}e^{a_{s}\tau^{B}_{s}}\simeq 2.12\times 10^{32}\, .
\end{align}
Thus, we do indeed get a large volume solution. And finally, we notice that using this minimum, we can find a flat direction for the other two large 4-cycle volumes $\tau^{B}_{1}$ and $\tau^{B}_{2}$.
\begin{align}
\tau^{B}_{1}=\frac{\left(\tau^{B}_{2}\right)^{3/2}-6 \tau^{B}_{s}\sqrt{\tau^{B}_{2}}+4 \sqrt{2} \left(\tau^{B}_{s}\right)^{3/2}+18 \mathcal{V}}{9 \sqrt{\tau^{B}_{2}}}\, .
\end{align}
This kind of feature will be particular interesting in the context of fiber modulus mediated inflation. Recall that $J^{B}_{2}$ is, in fact, a $K3$ fiber in this case. Work on these direct cosmological consequences is already being pursued and will be presented in future manuscripts.

\section{Results and Discussion}\label{sec:conc}

In this section we present the results of our search for toric Swiss cheese manifolds within the Kreuzer--Skarke database, for those polytopes with $h^{1,1} \leq 6$.
As indicated in Table~\ref{tab:swisscheese}, this represents $23,573$ reflexive polytopes, giving rise to $101,681$ unique Calabi-Yau threefolds, of which $100,368$ are favorable. The first stage of the analysis was to scan these $\sim 10^5$ favorable geometries, implementing the search strategy outlined in Section~\ref{sec:algorithm}. This was performed using resources at the Massachusetts Green High Performance Computing Center.
The computations were performed on dual Intel E5 2650 CPUs with 128GB of RAM per node, and the total time consumed for this stage in the analysis was $4,930$ core-hours.

\begin{table}[t!]
  \centering
    \begin{tabular}{|r|r||c|c|c|c|c|c|}
    \hline
    \multicolumn{2}{|c||}{\parbox[c][2em][c]{5cm}{\centering$\mathbf{h^{1,1}(X)}$}} & \textbf{1} & \textbf{2} & \textbf{3} & \textbf{4} & \textbf{5} & \textbf{6} \\\hline
    \hline
    \multicolumn{8}{|c|}{\parbox[c][2em][c]{15cm}{\centering\textbf{Polytope, Triangulations, and Geometries}}} \\
    \hline
    \multicolumn{2}{|r||}{\parbox[c][2em][c]{5cm}{\centering\textbf{\# of Polytopes}}} & 5     & 36    & 244   & 1197  & 4990  & 17101 \\
    \hline
    \multicolumn{2}{|r||}{\parbox[c][2em][c]{5cm}{\centering\textbf{\# of Triangulations}}} & 5     & 48    & 526   & 5348  & 57050 & 590085 \\
    \hline
    \multicolumn{2}{|r||}{\parbox[c][2em][c]{5cm}{\centering\textbf{\# of Geometries}}} & 5     & 39    & 306   & 2014  & 13635 & 85682 \\\hline
    \hline
    \multicolumn{8}{|c|}{\parbox[c][2em][c]{15cm}{\centering\textbf{Favorable Polytope, Triangulations, and Geometries}}} \\
    \hline
    \multicolumn{2}{|r||}{\parbox[c][3em][c]{5cm}{\centering\textbf{\# of Favorable Polytopes}}} & 5     & 36    & 243   & 1185  & 4897  & 16608 \\
    \hline
    \multicolumn{2}{|r||}{\parbox[c][3em][c]{5cm}{\centering\textbf{\# of Favorable Triangulations}}} & 5     & 48    & 525   & 5330  & 56714 & 584281 \\
    \hline
    \multicolumn{2}{|r||}{\parbox[c][3em][c]{5cm}{\centering\textbf{\# of Favorable Geometries}}} & 5     & 39    & 305   & 2000  & 13494 & 84525 \\\hline
    \hline
    \multicolumn{8}{|c|}{\parbox[c][2em][c]{15cm}{\centering\textbf{Toric Swiss Cheese Geometries}}} \\
    \hline
    \multicolumn{2}{|r||}{\parbox[c][4em][c]{5cm}{\centering\textbf{\% of Favorable Geometries Scanned for Toric Swiss Cheese}}} & 100   & 100   & 100   & 100   & 100   & 100 \\
    \hline
    \multirow{5}[10]{*}{\parbox[c][7em][c]{3.5cm}{\centering\textbf{\# of Toric Swiss Cheese Geometries (w/ Homogeneity Condition)}}} & \parbox[c][2em][c]{1.5cm}{\centering $\mathbf{N_{L}=1}$} & -     & 32 (22) & 86 (84) & 173 (171) & 603 (577) & 1304 (1137) \\
\cline{2-8}          & \parbox[c][2em][c]{1.5cm}{\centering $\mathbf{N_{L}=2}$} & -     & -     & 23 (23) & 17 (17) & 12 (10) & 17 (13) \\
\cline{2-8}          & \parbox[c][2em][c]{1.5cm}{\centering $\mathbf{N_{L}=3}$} & -     & -     & -     & 1 (1) & 0 (0) & 0 (0) \\
\cline{2-8}          & \parbox[c][2em][c]{1.5cm}{\centering $\mathbf{N_{L}=4}$} & -     & -     & -     & -     & 0 (0) & 0 (0) \\
\cline{2-8}          & \parbox[c][2em][c]{1.5cm}{\centering $\mathbf{N_{L}=5}$} & -     & -     & -     & -     & -     & 0 (0) \\\hline
    \hline
    \multicolumn{8}{|c|}{\parbox[c][2em][c]{15cm}{\centering\textbf{Explicit LVS Geometries}}} \\
    \hline
    \multicolumn{2}{|r||}{\parbox[c][4em][c]{5cm}{\centering\textbf{\% of Favorable Geometries Scanned for Explicitness}}} & 100   & 100   & 100   & 99.9  & 91.97 & 82.21 \\
    \hline
    \multicolumn{2}{|r||}{\parbox[c][3em][c]{5cm}{\centering\textbf{\# of Explicit Geometries}}} & 5     & 24    & 80    & 0     & 0     & 0 \\
    \hline
    \end{tabular}%
  \caption{Statistics for explicitness and toric Swiss cheese scans over favorable Calabi--Yau threefold geometries. For the Toric Swiss Cheese Geometries, the results in parentheses denote cases that also satisfy the homogeneity condition.
  The numbers are slightly different from~\cite{Altman:2014bfa} because of improved triangulation.
  As the scan for explicitness has not completed at the time of writing, future versions of this preprint will have updated results.}
  \label{tab:swisscheese}
\end{table}%

It is clear from these results that there is a scarcity of Calabi-Yau threefolds $X$, whose volumes can be made explicit at higher values of $h^{1,1}(X)$. Recall from Section \ref{sec:class} that this does not mean that we cannot ever write the 6-cycle volume of these manifolds in terms of 4-cycle volumes, but merely that to do so might involve a linear combination of arbitrary square roots, for which it is difficult to scan.

We also see from Table \ref{tab:swisscheese} that there are few toric Swiss cheese manifolds $X$ at higher $h^{1,1}(X)$ with many large 4-cycles. Unfortunately, when $N_{S}=h^{1,1}(X)-N_{L}>2$, it becomes difficult to stabilize the axion component of the complexified 4-cycle moduli \cite{Cicoli:2008va}. However, we see that there are still 18 cases in $h^{1,1}(X)=4$ for which the K\"ahler moduli can still be explicitly stabilized through the LVS. We demonstrate an example of this in Section \ref{sec:minim}. We note in Section \ref{sec:algorithm} that these results can be expanded by running this scan iteratively, each time with an arbitrary rotation of the toric intersection tensor. 

Finally, we notice that a large number of the toric Swiss cheese solutions satisfy the homogeneity condition of Section \ref{sec:homogeneity}. This condition ensures that the effective potential contains terms with the correct order in $\mathcal{V}^{-1}$ for a minimum to exist. It is not always necessary to achieve such a minimum when $N_{S}>1$, but it greatly simplifies the minimization procedure \cite{Cicoli:2008va}.

In total, $2,268$ toric Swiss cheese manifolds were identified, $2,055$ of which satisfied the homogeneity condition. These solutions are distributed as shown in the penultimate section of Table~\ref{tab:swisscheese}. While these numbers represent only a subset of all the possible Swiss cheese manifolds that may exist in this dataset, they represent an ample starting point for phenomenological investigation with nearly 160 explicit examples of guaranteed moduli stabilization using the techniques of \cite{Cicoli:2008va}. A concrete example of this was explored in Section \ref{sec:examples}.

More general techniques that do not rely on the ability to trivialize the unknown basis change (see Equation (\ref{eq:permute})), such as those employed in \cite{Gray:2012jy} are far more computationally expensive. Nevertheless, we expect such cases to form the majority of all manifolds which admit a large volume limit, particularly for higher values of $h^{1,1}$ and/or greater numbers of large cycles. A full analysis, extending the results of~\cite{Gray:2012jy}, is currently underway.

The $\mathcal{O}(2,000)$ cases with $h^{1,1}>4$ are, to our knowledge, unknown before now. Such cases were not approachable using the \texttt{PALP} software \cite{Braun:2012vh}, and required the redesigned techniques described in~\cite{Altman:2014bfa}. These high Picard number cases include $29$ cases with two or more large cycles. These are of particular interest for the possibility of identifying flat directions which may be relevant for inflationary cosmology~\cite{Conlon:2005jm,Cicoli:2008gp,Burgess:2016owb,Cicoli:2016chb}.

The second stage of the analysis involved studying the $100,368$ favorable Calabi-Yau geometries and identifying directly when a basis of divisor classes can be found for which the volume can be written explicitly in terms of 4-cycle volumes. Given the difficulty of studying the dynamics of K\"ahler moduli in the 4-dimensional effective supergravity theory when the Calabi-Yau volume cannot be put into a fully-explicit or (preferably) strong form, we also flag those cases for which a suitable basis exists, and provide the necessary rotation matrix ($\mathbf{T}^{C}$ in Section~\ref{sec:class}). We expect these results to be expanded significantly in the future as the more general case is completed. While this brute force attempt was performed over the entire dataset of Calabi-Yau threefolds, the identification of the components of the basis matrix involves solving a large system of polynomials using Groebner basis techniques. As such, this stage is computationally expensive, and both the CPU time and physical memory required to find a solution for any given example can be hard to predict. The results of the second stage in the analysis are found in the final section of Table~\ref{tab:swisscheese}, and we find a total of 109 usable cases in $h^{1,1}\leq 3$.

All $2,268$ toric Swiss cheese manifolds can be queried immediately via  the on-line database at \url{www.rossealtman.com}, allowing for quick access to this subset for the purposes of model building or further phenomenological study\footnote{For further convenience, we also flag whether or not our solutions of the $A$- and $B$-bases are integer-valued.}.
Given the difficulty of studying the dynamics of K\"ahler moduli in the 4-dimensional effective supergravity theory when the Calabi-Yau volume cannot be put into a fully-explicit or (preferably) strong form, we also flag those cases for which a suitable basis exists, and provide the necessary rotation matrix $\mathbf{T}^{C}$.
We expect these results to be expanded significantly in the future as the more general case is completed.

\section*{Acknowledgements}

We thank James Gray for discussions.
The work of RA and BDN is supported in part by the National Science Foundation, under grant PHY-1620575.
VJ is supported by the South African Research Chairs Initiative and the National Research Foundation.
YHH would like to thank the Science and Technology Facilities Council, UK, for grant ST/J00037X/1, the Chinese Ministry of Education, for a Chang-Jiang Chair Professorship at NanKai University as well as the City of Tian-Jin for a Qian-Ren Scholarship, and Merton College, Oxford, for her enduring support.

\begin{appendices}
\section{Large Volume Scenario Identities} \label{sec:appa}

The following two appendices provide a start-to-finish, self-contained derivation of all of the relevant expressions for moduli stabilization in the Large Volume Scenario, as well as key expressions from toric geometry which expedite the final result. While much of this material is present in the papers cited in the reference section of this work, we find that many of the intermediate steps are elided in the literature, and thus a more complete, pedagogical treatment is warranted.

For a given Calabi-Yau threefold $X$, the volume form is given by

\begin{equation}
\cV=\frac{1}{3!}\kappa_{ijk}t^{i}t^{j}t^{k}\, .
\end{equation}
From this, we can derive the volume of the $i^{\text{th}}$ 4-cycle divisor
\begin{align}
\tau_{a}&=\der{\cV}{t^{a}}=\frac{1}{3!}\kappa_{ijk}\left(\der{t^{i}}{t^{a}}t^{j}t^{k}+t^{i}\der{t^{j}}{t^{a}}t^{k}+t^{i}t^{j}\der{t^{k}}{t^{a}}\right)\notag\\
&=\frac{1}{3!}\kappa_{ijk}\left(\delta^{i}_{\; a}t^{j}t^{k}+t^{i}\delta^{j}_{\; a}t^{k}+t^{i}t^{j}\delta^{k}_{\; a}\right)\notag\\
&=\frac{1}{2}\kappa_{ajk}t^{j}t^{k}\, .
\end{align}
%
%
%
%
For ease of computation, we will also derive the following two relations involving derivatives of 2-cycle divisor volumes
\begin{align}\label{delta}
\delta^{a}_{\; b}&=\der{\tau_{b}}{\tau_{a}}=\der{}{\tau_{a}}\left(\frac{1}{2}\kappa_{bjk}t^{j}t^{k}\right)\notag\\
&=\frac{1}{2}\kappa_{bjk}\left(\der{t^{j}}{\tau_{a}}t^{k}+t^{j}\der{t^{k}}{\tau_{a}}\right)\notag\\
&=\kappa_{bjk}t^{k}\der{t^{j}}{\tau_{a}}\, ,
\end{align}
\begin{align}\label{derdelta}
&0=\der{\delta^{a}_{\; b}}{\tau_{c}}=\kappa_{bjk}\left(\der{t^{k}}{\tau_{c}}\der{t^{j}}{\tau_{a}}+t^{k}\derder{t^{j}}{\tau_{c}}{\tau_{a}}\right)\notag\\
\Rightarrow\;\;\;&\kappa_{bjk}t^{k}\derder{t^{j}}{\tau_{c}}{\tau_{a}}=-\kappa_{bjk}\der{t^{k}}{\tau_{c}}\der{t^{j}}{\tau_{a}}
, .
\end{align}
Using Equation (\ref{delta}), we now compute the first derivative of the volume form
\begin{align}
\cV^{a}&=\der{\cV}{\tau_{a}}=\frac{1}{3!}\kappa_{ijk}\left(\der{t^{i}}{\tau_{a}}t^{j}t^{k}+t^{i}\der{t^{j}}{\tau_{a}}t^{k}+t^{i}t^{j}\der{t^{k}}{\tau_{a}}\right)\notag\\
&=\frac{1}{2}\der{t^{i}}{\tau_{a}}\kappa_{ijk}t^{j}t^{k}=\frac{1}{2}\delta^{\; a}_{k}t^{k}=\frac{1}{2}t^{a}\, ,
\end{align}
and using Equation (\ref{derdelta}), we compute the second derivative
\begin{align}
\cV^{ab}&=\der{\cV^{a}}{\tau_{b}}=\frac{1}{2}\kappa_{ijk}\left(\derder{t^{i}}{\tau_{b}}{\tau_{a}}t^{j}t^{k}+\der{t^{i}}{\tau_{a}}\der{t^{j}}{\tau_{b}}t^{k}+\der{t^{i}}{\tau_{a}}t^{j}\der{t^{k}}{\tau_{b}}\right)\notag\\
&=\frac{1}{2}\kappa_{ijk}\left(-\der{t^{k}}{\tau_{b}}\der{t^{i}}{\tau_{a}}t^{j}+\der{t^{i}}{\tau_{a}}\der{t^{j}}{\tau_{b}}t^{k}+\der{t^{i}}{\tau_{a}}t^{j}\der{t^{k}}{\tau_{b}}\right)\notag\\
&=\frac{1}{2}\delta^{b}_{\; i}\der{t^{i}}{\tau_{a}}=\frac{1}{2}\der{t^{b}}{\tau_{a}}\, .
\end{align}
Again using Equation (\ref{delta}), we can then derive the inverse of the second derivative of the volume form
\begin{align}\label{eq:AVinv}
&\delta^{a}_{\; b}=\kappa_{bjk}t^{k}\der{t^{j}}{\tau_{a}}=2\kappa_{bjk}t^{k}\cV^{aj}\notag\\
\Rightarrow\;\;\;&\icV_{jb}=2\kappa_{jbk}t^{k}\, .
\end{align}
We now derive the following three relations, which will be useful in computing the inverse K\"ahler metric
\begin{align}\label{vabivbc}
\cV^{ab}\icV_{bc}&=\left(\frac{1}{2}\der{t^{b}}{\tau_{a}}\right)\left(2\kappa_{bck}t^{k}\right)=\delta^{a}_{\; c}\, ,
\end{align}
\begin{align}\label{vaivab}
\cV^{a}\icV_{ab}&=\left(\frac{1}{2}t^{a}\right)\left(2\kappa_{abk}t^{k}\right)=2\tau_{b}\, ,
\end{align}
\begin{align}\label{vaivabvb}
\cV^{a}\icV_{ab}\cV^{b}&=\left(\frac{1}{2}t^{a}\right)\left(2\kappa_{abk}t^{k}\right)\left(\frac{1}{2}t^{b}\right)=3\cV\, .
\end{align}

\section{K\"ahler Moduli Stabilization}\label{sec:appb}

The bosonic field content in the string frame is as follows
\begin{itemize}
\item R-R sector:
A 0-form potential $C_{0}$, a 2-form potential $C_{2}$, and a 4-form potential $C_{4}$.
\item NS-NS sector:
The 0-form dilaton $\phi$, the 2-form 10D graviton $g_{\mu\nu}$, and the antisymmetric Kalb-Ramond 2-form $B_{2}$.
\item Scalar moduli:
K\"ahler moduli ($\tau_{i}$) and complex structure moduli ($U_{i}$).
\end{itemize}
A constant variation in the dilaton can be shown to produce a corresponding variation in the string coupling according to $\frac{\delta g_{s}}{g_{s}}=\delta\phi$, so that we may express the coupling as $g_{s}\sim e^{\phi}$. Furthermore, in the Einstein frame, the 10D string-frame graviton is rescaled by $g^{(s)}_{\mu\nu}\rightarrow g^{(E)}_{\mu\nu}=g_{s}^{-1/2}g^{(s)}_{\mu\nu}$. This results in the rescaling of each 2-cycle volume as $t^{i}\rightarrow \hatt^{i}=g_{s}^{-1/2}t^{i}$, and therefore $\tau_{i}\rightarrow \hattau_{i}=g_{s}^{-1}\tau_{i}$ and $\cV\rightarrow \hatcV=g_{s}^{-3/2}\cV$.
It is convenient to make the following field redefinitions
\begin{itemize}
\item
The axion-dilaton $S=g_{s}^{-1}+iC_{0}$, which corresponds to the complex structure of the elliptic fiber in the F-theory generalization.
\item
The complexified K\"ahler moduli $T_{i}=\int\limits_{J_{i}}{\left(J\wedge J+iC_{4}\right)}=\hattau_{i}+ib_{i}$.
\end{itemize}
In this section, we will show that at tree level, the scalar potential of the 4D effective supergravity theory exhibits a ``no-scale'' structure, at which only the axion-dilaton and complex structure moduli are stabilized. We further show that in order to break the ``no-scale'' structure and stabilize the volume modulus, we must include the leading $\alpha '$ correction to the volume in the K\"ahler potential. And finally, we show that to stabilize the remaining K\"ahler blowup modes, we must consider non-perturbative corrections to the superpotential resulting from the structure on the blowup cycles. In the end, we will write down the corrected scalar potential
\begin{equation}
\begin{array}{ccccccc}
V&=&V_{\text{tree}}&+&V_{\alpha '}&+&V_{\text{non-perturbative}}
\end{array}
\end{equation}
\noindent which can be minimized to stabilize the K\"ahler blowup moduli, as well as the volume modulus, which gets fixed exponentially large with respect to the blowup moduli. The presence of a flat direction in the moduli space at this minimum leaves the door open for K\"ahler blowup moduli-mediated inflation. We will discuss this in the next section.

In the absence of flat directions, the blowup moduli are all stabilized at small values. However, this still leaves any non-blowup K\"ahler moduli (which correspond instead to fibration modes and typically have large values) unstabilized. This ``extended no-scale'' structure can only be broken by adding subleading string loop corrections. The fibration moduli can then be stabilized by minimizing the further corrected potential. 

\subsection{$V_{\text{tree}}$}

At tree level in $\alpha '$, the K\"ahler potential for $X$ can be expressed in the separated form\footnote{Technically, the Einstein frame volume $\hatcV=g_{s}^{-3/2}\cV$ depends on the axion-dilaton $S$ through $g_{s}\sim\frac{S+\bar{S}}{2}$. However, we can disregard this, since we will be differentiating with respect to the Einstein frame K\"ahler moduli $\hattau$, which have a complementary dependence on $S$.}
\begin{equation}
\begin{array}{ccccccc}
\cK&=&\cK_{S}&+&\cK_{T}&+&\cK_{U}\\
 &=&-\text{ln}\left(S+\bar{S}\right)&+&-2\, \text{ln}\left(\hatcV\right)&+&-\text{ln}\left(-i\int\limits_{X}{\Omega\wedge\bar{\Omega}}\right)\, ,
\end{array}
\end{equation}
\noindent where $\Omega$ is the unique, holomorphic (3,0)-form on $X$ which contains the dependence on the complex structure moduli $U_{i},\; i=1,...,h^{2,1}$. If we turn on the non-trivial RR and NS-NS gauge fluxes $F_{3}=dC_{2}$ and $H_{3}=dB_{2}$, and construct the complexified flux $G_{3}=F_{3}+iSH_{3}$, then the superpotential can be written as
\begin{align}
W=W_{\text{Gukov-Vafa-Witten}}=\int\limits_{X}{\Omega\wedge G_{3}}\, .
\end{align}
In order to obtain a warped 10D background, the Bianchi identity constrains $G_{3}$ to be imaginary self-dual (i.e. $*_{6}G_{3}=iG_{3}$). It can be shown that this is equivalent to the constraints on the GVW superpotential
\begin{align}\label{eq:csstable}
D^{A}W=0,\;\;\Phi_{A}\in\{S,U_{i}\}\, .
\end{align}
The full scalar potential in the 4D effective supergravity theory then has the form
\begin{align}
V=e^{\cK}\left[\icK_{a\bar{b}}D^{a}WD^{\bar{b}}W-3|W|^{2}\right],\;\; \Phi_{a},\Phi_{b}\in\{S,T_{1},...,T_{h^{1,1}},U_{1},...,U_{h^{2,1}}\}
\end{align}
\noindent where the inverse K\"ahler metric is defined as
\begin{align*}
\icK_{a\bar{b}}\equiv \left(\frac{\partial\cK}{\partial\Phi_{a}\bar{\partial\Phi_{b}}}\right)^{-1}
\end{align*}
\noindent and the gauge connection for the covariant derivative is given by $\partial_{a}\cK\equiv\frac{\partial\cK}{\partial\Phi_{a}}$ so that
\begin{align*}
D^{a}W=\partial^{a}W+W\partial^{a}\cK\, .
\end{align*}
Because $\cK$ is separated into $\cK_{S}$, $\cK_{T}$, and $\cK_{U}$, the inverse K\"ahler metric has a block diagonal form
\begin{align}
\icK_{a\bar{b}}=\left(\begin{array}{ccc}
\left(\cK_{S}^{-1}\right)&0&0\\
0&\icKK_{i\bar{j}}&0\\
0&0&\left(\cK_{U}^{-1}\right)_{A\bar{B}}\\
\end{array}\right)
\end{align}
\noindent where $\Phi_{i},\Phi_{j}\in\{T_{1},...,T_{h^{1,1}}\}$ and $\Phi_{A},\Phi_{B}\in\{U_{1},...,U_{h^{2,1}}\}$. Then, the scalar potential separates
\begin{equation}\label{eq:spotdecomp}
\begin{array}{ccccccccc}
V&=&V_{S}&+&V_{T}&+&V_{U}&+&-3e^{K}|W|^2\\
 &=&0&+&V_{T}&+&0&+&-3e^{K}|W|^2
\end{array}\, .
\end{equation}
\noindent where the second equality follows from the constraints in Equation (\ref{eq:csstable}). Then, focusing on $V_{T}$, we have
\begin{align}\label{eq:VKahler}
V_{T}=e^{\cK}\left[\icKK_{i\bar{j}}D^{i}WD^{\bar{j}}W\right]
\end{align}
The first derivative of $K_{T}$ is given by
\begin{align}
\cK_{T}^{i}&=\partial^{i}\cK_{T}=\frac{\partial}{\partial T_{i}}\cK_{T}=-2\hatcV^{-1}\frac{\partial\hatcV}{\partial T_{i}}=-2\hatcV^{-1}\frac{1}{2}\left(\frac{\partial\hatcV}{\partial \hattau_{i}}-i\frac{\partial\hatcV}{\partial b_{i}}\right)\notag\\
&=-\hatcV^{-1}\hatcV^{i}\, .
\end{align}
Then, the second derivative is given by
\begin{align}
\cK_{T}^{i\bar{j}}&=\partial^{i}\partial^{\bar{j}}\cK_{T}=-\left(\frac{\partial}{\partial T_{i}}\hatcV^{-1}\right)\hatcV^{j}-\hatcV^{-1}\left(\frac{\partial}{\partial T_{i}}\hatcV^{j}\right)\notag\\
&=\hatcV^{-2}\frac{1}{2}\left(\frac{\partial\hatcV}{\partial \hattau_{i}}-i\frac{\partial\hatcV}{\partial b_{i}}\right)\hatcV^{j}-\hatcV^{-1}\frac{1}{2}\left(\frac{\partial\hatcV^{j}}{\partial \hattau_{i}}-i\frac{\partial\hatcV^{j}}{\partial b_{i}}\right)\notag\\
&=\frac{1}{2\hatcV}\left(\frac{1}{\hatcV}\hatcV^{i}\hatcV^{j}-\hatcV^{ij}\right)\, .
\end{align}
Now, we assume that $\icKK_{i\bar{j}}$ is of the form
\begin{align}\label{eq:invKassum}
\icKK_{i\bar{j}}&=u\hatcV\ihatcV_{ij}+v\hatcV^{k}\ihatcV_{ki}\hatcV^{l}\ihatcV_{lj}+\mathcal{O}\left(\text{higher order in }\hatcV^{-1}\right)\, .
\end{align}
Then, we have
\begin{align}
\cK_{T}^{i\bar{j}}\icKK_{\bar{j}k}&=\frac{1}{2\hatcV}\left[\frac{1}{\hatcV}\hatcV^{i}\hatcV^{j}-\hatcV^{ij}\right]\left[u\hatcV\ihatcV_{jk}+v\hatcV^{l}\ihatcV_{lj}\hatcV^{m}\ihatcV_{mk}\right]\notag\\
&=\frac{1}{2\hatcV}\left[u\hatcV^{i}\hatcV^{j}\ihatcV_{jk}+\frac{v}{\hatcV}\hatcV^{i}\hatcV^{l}\ihatcV_{lj}\hatcV^{j}\hatcV^{m}\ihatcV_{mk}-u\hatcV\hatcV^{ij}\ihatcV_{jk}\right.\notag\\
&\hspace{1.2cm}\left.-v\hatcV^{ij}\ihatcV_{jl}\hatcV^{l}\hatcV^{m}\ihatcV_{mk}\right]\notag\\
&=\frac{1}{2\hatcV}\left[u\hatcV^{i}\hatcV^{j}\ihatcV_{jk}+\frac{v}{\hatcV}\hatcV^{i}\left(3\hatcV\right)\hatcV^{m}\ihatcV_{mk}-u\hatcV\delta^{i}_{\; k}-v\delta^{i}_{\; l}\hatcV^{l}\hatcV^{m}\ihatcV_{mk}\right]\notag\\
&=\frac{1}{2\hatcV}\left[u\hatcV^{i}\hatcV^{j}\ihatcV_{jk}+3v\hatcV^{i}\hatcV^{j}\ihatcV_{jk}-u\hatcV\delta^{i}_{\; k}-v\hatcV^{i}\hatcV^{j}\ihatcV_{jk}\right]\notag\\
&=\frac{1}{2\hatcV}\left[\left(u+2v\right)\hatcV^{i}\hatcV^{j}\ihatcV_{jk}-u\hatcV\delta^{i}_{\; k}\right]\notag\\
&=\left(\frac{u}{2}+v\right)\frac{1}{\hatcV}\left(\frac{1}{2}\hatt^{i}\right)\left(2\hattau_{k}\right)-\frac{u}{2}\delta^{i}_{\; k}\notag\\
&=\left(\frac{u}{2}+v\right)\frac{\hatt^{i}\hattau_{k}}{\hatcV}-\frac{u}{2}\delta^{i}_{\; k}\, .
\end{align}
In order for this result to be consistent and general, we must have
\begin{align}
u=-2\hspace{1cm}\text{and}\hspace{1cm}v=1
\end{align}
\noindent so that
\begin{align}
\icKK_{i\bar{j}}&=-2\hatcV\ihatcV_{ij}+\hatcV^{k}\ihatcV_{ki}\hatcV^{l}\ihatcV_{lj}+\mathcal{O}\left(\text{higher order in }\hatcV^{-1}\right)\, .
\end{align}
We also know that since $W=W_{\text{GVW}}$ is independent of the K\"ahler moduli, we can write
\begin{align}
D^{i}W&=W\partial^{i}\cK_{T}=-W\hatcV^{-1}\hatcV^{i}\, .
\end{align}
Then, expanding Equation (\ref{eq:VKahler}), we have
\begin{align}
V_{T}&=e^{\cK}\left[\icKK_{i\bar{j}}D^{i}WD^{\bar{j}}W\right]\notag\\
&=e^{\cK}|W|^{2}\left[-2\hatcV\ihatcV_{ij}+\hatcV^{k}\ihatcV_{ki}\hatcV^{l}\ihatcV_{lj}\right]\hatcV^{-2}\hatcV^{i}\hatcV^{j}\notag\\
&=e^{\cK}|W|^{2}\hatcV^{-2}\left[-2\hatcV\hatcV^{i}\ihatcV_{ij}\hatcV^{j}+\hatcV^{k}\ihatcV_{ki}\hatcV^{i}\hatcV^{l}\ihatcV_{lj}\hatcV^{j}\right]\notag\\
&=e^{\cK}|W|^{2}\hatcV^{-2}\left[-2\hatcV\left(3\hatcV\right)+\left(3\hatcV\right)\left(3\hatcV\right)\right]\notag\\
&=3e^{\cK}|W|^{2}\, ,
\end{align}
\noindent and the scalar potential in Equation (\ref{eq:spotdecomp}) reduces to the ``no-scale'' form with
\begin{align}
V_{\text{tree}}=0\, .
\end{align}
Thus, at tree level, this potential cannot be minimized, and so both the volume and the individual K\"ahler moduli are left unstabilized.

\subsection{$V_{\alpha '}$}

In order to stabilize the volume $\hatcV$, we must break the ``no-scale'' structure. We do this by considering the leading $(\alpha ')^{3}$ correction to $\cV$, given by $\frac{\xi}{2}$, where\footnote{$\chi (X)$ is the Euler characteristic of $X$, and $\zeta (3)\approx 1.20206$ is the Riemann zeta function, evaluated at 3.} $\xi =-\frac{\chi (X)\zeta (3)}{2}$. With this correction, the Einstein frame volume becomes
\begin{align}
\hatcV& =g_{s}^{-3/2}\cV\notag\\
&\rightarrow\;\; g_{s}^{-3/2}\left(\cV + \frac{\xi}{2}\right)=\hatcV + \frac{\xi}{2}\left(\frac{S+\bar{S}}{2}\right)^{3/2}\, .
\end{align}
Then, the K\"ahler potential for $X$ can be expressed in a partially separated form as
\begin{equation}\label{eq:aprimesplit}
\begin{array}{ccccc}
\cK&=&\cK_{S/T}&+&\cK_{U}\\
 &=&-\text{ln}\left(S+\bar{S}\right)-2\, \text{ln}\left(\hatcV +\frac{\xi}{2}\left(\frac{S+\bar{S}}{2}\right)^{3/2}\right)&+&-\text{ln}\left(-i\int\limits_{X}{\Omega\wedge\bar{\Omega}}\right)\, .
\end{array}
\end{equation}
For ease of notation, we also define
\begin{align}
\cK_{T}=-2\, \text{ln}\left(\hatcV +\frac{\xi}{2}\left(\frac{S+\bar{S}}{2}\right)^{3/2}\right)=-2\, \text{ln}\left(\hatcV +\frac{\hat{\xi}}{2}\right)\, .
\end{align}
This allows us to write $K^{i\bar{j}}=\partial^{i}\partial^{\bar{j}}K$ in block diagonal form
\begin{align}
\cK^{\bar{a}b}&=\left(\begin{array}{c|cc}
\left(\cK_{S/T}\right)^{00}&\left(\cK_{S/T}^{T}\right)^{0j}&0\\
\hline\\
\left(K_{S/T}\right)^{\bar{i0}}&\left(\cK_{T}\right)^{\bar{i}j}&0\\
\\
0&0&\left(\cK_{U}\right)^{\bar{A}B}
\end{array}\right)\notag\\
\notag\\
&=\left(\begin{array}{c|c}
a&\mathbf{b}^{T}\\
\hline
\mathbf{b}&\mathbf{C}
\end{array}\right)\, ,
\end{align}
where $\Phi_{i},\Phi_{j}\in\{T_{1},...,T_{h^{1,1}}\}$ and $\Phi_{A},\Phi_{B}\in\{U_{1},...,U_{h^{2,1}}\}$. The inverse is given by
\begin{align}\label{eq:geninverse}
\icK_{a\bar{b}}=\frac{1}{d}\left(\begin{array}{c|c}
1&-\mathbf{b}^{T}\mathbf{C}^{-1}\\
\hline
-\mathbf{C}^{-1}\mathbf{b}&d\mathbf{C}^{-1}+\mathbf{C}^{-1}\mathbf{b}\mathbf{b}^{T}\mathbf{C}^{-1}
\end{array}\right)\, ,
\end{align}
\noindent where $d=a-\mathbf{b}^{T}\mathbf{C}^{-1}\mathbf{b}$. In addition, our matrix $\mathbf{C}$ is diagonal, so
\begin{align}
\mathbf{C}^{-1}&=\left(\begin{array}{cc}
\left(\cK^{-1}_{T}\right)_{i\bar{j}}&0\\
0&\left(\cK^{-1}_{U}\right)_{A\bar{B}}
\end{array}\right)
\end{align}
\noindent and Equation (\ref{eq:geninverse}) reduces to
\begin{align}\label{eq:fullinv}
&\icK_{a\bar{b}}=\frac{1}{d}\times\notag\\
&\left(\begin{array}{c|cc}
1&-\left(\cK_{S/T}^{T}\right)^{0k}\left(\cK^{-1}_{T}\right)_{k\bar{j}}&0\\
\hline\\
-\left(\cK^{-1}_{T}\right)_{i\bar{k}}\left(\cK_{S/T}\right)^{\bar{k}0}&d\left(\cK^{-1}_{T}\right)_{i\bar{j}}+\left(\cK^{-1}_{T}\right)_{i\bar{k}}\left(\cK^{T}_{S/T}\right)^{\bar{k}0}\left(\cK_{S/T}\right)^{0l}\left(\cK^{-1}_{T}\right)_{l\bar{j}}&0\\
\\
0&0&d\left(\cK^{-1}_{U}\right)_{A\bar{B}}
\end{array}\right)\, ,
\end{align}
\noindent where $d=\left(\cK_{S/T}\right)^{00}-\left(\cK_{S/T}^{T}\right)^{0i}\left(\cK^{-1}_{T}\right)_{i\bar{j}}\left(\cK_{S/T}\right)^{\bar{j}0}$.

The full scalar potential in the 4D effective supergravity theory again has the form
\begin{align}
V=e^{\cK}\left[\icK_{a\bar{b}}D^{a}WD^{\bar{b}}W-3|W|^{2}\right],\;\; \Phi_{a},\Phi_{b}\in\{S,T_{1},...,T_{h^{1,1}},U_{1},...,U_{h^{2,1}}\}\, .
\end{align}
Due to the superpotential constraints (see Equation (\ref{eq:csstable})) that stabilize the axion-dilaton and complex structure moduli at tree level, any terms proportional to $D^{S}W$ or $D^{U_{i}}W$ vanish, and we need only consider the center block of the Equation (\ref{eq:fullinv}). This gives us the corrected inverse K\"ahler metric\cite{Bobkov:2004cy,Becker:2002nn}
\begin{align}\label{eq:BKinvT}
\left(\tilde{\cK}_{T}^{-1}\right)_{i\bar{j}}=\icKK_{i\bar{j}}+\frac{1}{d}\icKK_{i\bar{k}}\left(\cK^{T}_{S/T}\right)^{\bar{k}0}\left(\cK_{S/T}\right)^{0l}\icKK_{l\bar{j}}
\end{align}
\noindent where, from Equation (\ref{eq:aprimesplit}) and Equation (\ref{eq:invKassum}) with $\hatcV\rightarrow\hatcV+\frac{\hat{\xi}}{2}$ and $\hat{\xi}=g_{s}^{-3/2}\xi$, we find that
\begin{align}
\icKK_{i\bar{j}}&=-\left(2\hatcV+\hat{\xi}\right)\ihatcV_{ij}+2\left(\frac{2\hatcV+\hat{\xi}}{4\hatcV-\hat{\xi}}\right)\hatcV^{k}\ihatcV_{ki}\hatcV^{l}\ihatcV_{lj}\, , \\
\notag\\
\left(\cK_{S/T}\right)^{0i}&=\frac{3\hat{\xi}g_{s}}{4\left(2\hatcV+\hat{\xi}\right)^2}\hatcV^{i}\, ,\\
\notag\\
\left(\cK_{S/T}\right)^{i}&=-\frac{2}{2\hatcV+\hat{\xi}}\hatcV^{i}\, , \\
\notag\\
\text{and}\;\;\;\;\;\; d&=\frac{g_{s}^{2}\left(\hatcV-\hat{\xi}\right)}{4\left(4\hatcV-\hat{\xi}\right)}\, .
\end{align}
The scalar potential then takes the form
\begin{align}\label{eq:scalargenaprime}
V=e^{\cK}\left[\left(\tilde{\cK}^{-1}_{T}\right)_{i\bar{j}}D^{i}WD^{\bar{j}}W-3|W|^{2}\right]\, .
\end{align}
Recall that the superpotential is independent of the K\"ahler moduli, so that $D^{i}W=W\partial^{i}\cK=W\left(\cK_{S/T}\right)^{i}$. Then, the scalar potential reduces to
\begin{align}
\begin{array}{ccccc}
V&=&V_{\text{tree}}&+&V_{\alpha '}\\
\\
&=&0&+&3e^{\cK}|W|^{2}\hat{\xi}\frac{\left(\hat{\xi}^{2}+7\hat{\xi}\hatcV+\hatcV^{2}\right)}{\left(\hatcV-\hat{\xi}\right)\left(2\hatcV+\hat{\xi}\right)^{2}}\, .
\end{array}
\end{align}
We can now find a stable minimum for the volume modulus.

\subsection{$V_{\text{non-perturbative}}$}

In the previous subsection, we were able to use the leading $\alpha '$ correction to the volume in order to break the ``no-scale'' structure of the potential and stabilize the volume modulus, but this still did not give us a mechanism for stabilizing the remaining K\"ahler moduli. In order to find such a mechanism, we must further consider the effect of non-perturbative features on the superpotential. The superpotential then takes the form

\begin{equation}
\begin{array}{ccccc}
W&=&W_{\text{GVW}}&+&W_{\text{non-perturbative}}\\
 &=&\int\limits_{X}{\Omega\wedge G_{3}}&+&A_{i}e^{-\hat{a}_{i}T_{i}}
\end{array}
\end{equation}
\noindent where the scalar constants $\hat{a}_{i}$ depend on non-perturbative effects such as D brane instantons ($\hat{a}_{i}=\frac{2\pi}{g_{s}}$) or gaugino condensation ($\hat{a}_{i}=\frac{2\pi}{g_{s}N}$) on the corresponding 4-cycle $J_{i}\in H_{4}(X;\mathbb{Z})$, and the complex constants $A_{i}$ encode threshold effects depending implicitly on the complex structure and D3 brane positions.

The scalar potential still takes the form in Equation (\ref{eq:scalargenaprime}), except that
\begin{align}
D^{i}W=\partial^{i}W+W\partial^{i}\cK=-\hat{a}_{i}A_{i}e^{-\hat{a}_{i}T_{i}}+W\left(\cK_{\text{ad/K}}\right)^{i}\, .
\end{align}
Plugging this in, we find that
\begin{align}
V=e^{\cK}&\left[\left(\tilde{\cK}^{-1}_{T}\right)_{i\bar{j}}\hat{a}_{i}\hat{a}_{j}A_{i}\bar{A}_{j}e^{-\left(\hat{a}_{i}T_{i}+\hat{a}_{j}\bar{T}_{j}\right)}\right.\notag\\
&\left.-\left(\tilde{\cK}^{-1}_{T}\right)_{i\bar{j}}\left(\hat{a}_{i}A_{i}\bar{W}e^{-\hat{a}_{i}T_{i}}\left(\cK_{S/T}\right)^{\bar{j}}+\left(\cK_{S/T}\right)^{i}\hat{a}_{j}\bar{A}_{j}We^{-\hat{a}_{j}\bar{T}_{j}}\right)\right.\notag\\
&+\left.\left(\tilde{\cK}^{-1}_{T}\right)_{i\bar{j}}|W|^{2}\left(\cK_{S/T}\right)^{i}\left(\cK_{S/T}\right)^{\bar{j}}-3|W|^{2}\right]\, ,
\end{align}
\noindent where the last line is just $V_{\alpha '}$. This then reduces to
\begin{align}
\begin{array}{ccccccccc}
V&=&V_{\text{tree}}&+&V_{\text{np1}}&+&V_{\text{np2}}&+&V_{\alpha '}\\
\end{array}
\end{align}
\noindent with
\begin{align}
V_{\text{np1}}&=e^{\cK}\left(\tilde{\cK}^{-1}_{T}\right)_{i\bar{j}}\hat{a}_{i}\hat{a}_{j}A_{i}\bar{A}_{j}e^{-\left(\hat{a}_{i}T_{i}+\hat{a}_{j}\bar{T}_{j}\right)}\notag\\
&=e^{\cK}\hat{a}_{i}\hat{a}_{j}A_{i}\bar{A}_{j}e^{-\left(\hat{a}_{i}T_{i}+\hat{a}_{j}\bar{T}_{j}\right)}\left(2\hatcV+\hat{\xi}\right)\left[-\ihatcV_{ij}+\left(\frac{2}{4\hatcV-\hat{\xi}}\right)\hatcV^{k}\ihatcV_{ki}\hatcV^{l}\ihatcV_{lj}\right]\notag\\
&=2e^{\cK}\hat{a}_{i}\hat{a}_{j}\left\lvert A_{i}A_{j}\right\rvert e^{-\left(\hat{a}_{i}\hattau_{i}+\hat{a}_{j}\hattau_{j}\right)}e^{i\left(\theta_{i}-\theta_{j}-\hat{a}_{i}b_{i}+\hat{a}_{j}b_{j}\right)}\left(2\hatcV+\hat{\xi}\right)\left[-\kappa_{ijk}\hatt^{k}+\left(\frac{4}{4\hatcV-\hat{\xi}}\right)\hattau_{i}\hattau_{j}\right]
\end{align}
\noindent and
\begin{align}
V_{\text{np2}}&=-e^{\cK}\left(\tilde{\cK}^{-1}_{T}\right)_{i\bar{j}}\left(\hat{a}_{i}A_{i}\bar{W}e^{-\hat{a}_{i}T_{i}}\left(\cK_{S/T}\right)^{\bar{j}}+\left(\cK_{S/T}\right)^{i}\hat{a}_{j}\bar{A}_{j}We^{-\hat{a}_{j}\bar{T}_{j}}\right)\notag\\
&=e^{\cK}\hat{a}_{i}\frac{4\hat{\xi}^{2}+\hat{\xi}\hatcV +4\hatcV^{2}}{\left(\hatcV -\hat{\xi}\right)\left(\hat{\xi}+2\hatcV\right)}\left(A_{i}\hattau_{i}\bar{W}e^{-\hat{a}_{i}T_{i}}+\bar{A}_{i}\hattau_{i}We^{-\hat{a}_{i}\bar{T}_{i}}\right)\notag\\
&=2e^{\cK}\hat{a}_{i}\left\lvert A_{i}W\right\rvert\hattau_{i}e^{-\hat{a}_{i}\hattau_{i}}\frac{4\hat{\xi}^{2}+\hat{\xi}\hatcV +4\hatcV^{2}}{\left(\hatcV -\hat{\xi}\right)\left(2\hatcV+\hat{\xi}\right)}\text{cos}\left(\theta_{i}-\phi -\hat{a}_{i}b_{i}\right)\, ,
\end{align}
\noindent where $A_{i}=\left\lvert A_{i}\right\rvert e^{i\theta_{i}}$ and $W=|W|e^{i\phi}$.

The axionic part $b_{i}$ of the complexified K\"ahler moduli can be decoupled and stabilized independently. The result, however, depends heavily on the topology of $X$ as encoded in the triple intersection tensor $\kappa_{ijk}$, and therefore its complexity also scales rapidly with increasing numbers of blowup moduli. In their appendix, the authors of \cite{Cicoli:2008va} did an excellent job of classifying the resulting axion-stabilized scalar potential for various forms of $\kappa_{ijk}$ in the large volume limit for up to two blowup moduli, and the reader is encouraged to refer there for more detail.

In this appendix, we consider only the ``Swiss cheese'' case in which the small 4-cycle blowup moduli can be explicitly separated from the large moduli which control the volume and flbration structure. In addition, for the sake of simplicity, we turn our attention only to cases with one small blowup modulus $\hattau_{s}$, while the rest are sent large. In this case, it is a relatively simple matter to stabilize the single axion $b_{s}$, as its contribution cancels in $V_{\text{np1}}$. We find that
\begin{align}
V_{\text{np1}}&=2e^{\cK}\hat{a}_{s}^{2}\left\lvert A_{s}\right\rvert^{2}e^{-2\hat{a}_{s}\hattau_{s}}\left(2\hatcV+\hat{\xi}\right)\left(-\kappa_{ssi}\hatt^{i}+\left(\frac{4}{4\hatcV-\hat{\xi}}\right)\hattau_{s}\hattau_{s}\right)\\
\notag\\
V_{\text{np2}}&=-2e^{\cK}\hat{a}_{s}\left\lvert A_{s}W\right\rvert e^{-\hat{a}_{s}\hattau_{s}}\hattau_{s}\frac{4\hat{\xi}^{2}+\hat{\xi}\hatcV +4\hatcV^{2}}{\left(\hatcV -\hat{\xi}\right)\left(\hat{\xi}+2\hatcV\right)}\notag\\
&=-2e^{\cK}\hat{a}_{s}\left\lvert A_{s}\right\rvert e^{-\hat{a}_{s}\hattau_{s}}\left(\left\lvert W_{\text{GVW}}\right\rvert-\left\lvert A_{s}\right\rvert e^{-\hat{a}_{s}\hattau_{s}}\right)\hattau_{s}\frac{4\hat{\xi}^{2}+\hat{\xi}\hatcV +4\hatcV^{2}}{\left(\hatcV -\hat{\xi}\right)\left(2\hatcV+\hat{\xi}\right)}\, .
\end{align}
Furthermore, \cite{Cicoli:2008va} shows that in the case of a single small blowup modulus, there will only be a large volume AdS minimum when an additional so-called ``homogeneity condition''
\begin{align}\label{eq:homogeneity}
\kappa_{ssi}\hatt^{i}\simeq -c\sqrt{\hattau_{s}},\;\;\;c>0
\end{align}
\noindent is satisfied\footnote{The minus sign in Equation (\ref{eq:homogeneity}) originates from the fact that inside the K\"ahler cone $\int\limits_{C^{i}}{J}>0$, the K\"ahler metric must be positive definite.}. Then, in the large volume limit we have $e^{\cK}\underset{\hatcV\rightarrow\infty}{=}\frac{g_{s}e^{\cK_{\text{cs}}}}{2\hatcV^{2}}$, and to leading order in each term
\begin{align}
V_{\text{np1}}&\underset{\hatcV\rightarrow\infty}{=}g_{s}e^{\cK_{\text{cs}}}\frac{2c\hat{a}_{s}^{2}\left\lvert A_{s}\right\rvert^{2}e^{-2\hat{a}_{s}\hattau_{s}}\sqrt{\hattau_{s}}}{\hatcV}\, , \\
\notag\\
V_{\text{np2}}&\underset{\hatcV\rightarrow\infty}{=}-g_{s}e^{\cK_{\text{cs}}}\frac{2\hat{a}_{s}\left\lvert A_{s}W_{\text{GVW}}\right\rvert e^{-\hat{a}_{s}\hattau_{s}}\hattau_{s}}{\hatcV^{2}}\, ,\\
\notag\\
V_{\alpha '}&\underset{\hatcV\rightarrow\infty}{=}g_{s}e^{\cK_{\text{cs}}}\frac{3\left\lvert W_{\text{GVW}}\right\rvert^{2}\hat{\xi}}{8\hatcV^{3}}\, ,
\end{align}
\noindent and we obtain the full potential
\begin{align}\label{eq:standpot}
V\left(\hattau_{s},\hatcV\right)&=V_{\text{tree}}+V_{\text{np1}}+V_{\text{np2}}+V_{\alpha '}\notag\\
&=g_{s}e^{\cK_{\text{cs}}}\left[\frac{2c\hat{a}_{s}^{2}\left\lvert A_{s}\right\rvert^{2}e^{-2\hat{a}_{s}\hattau_{s}}\sqrt{\hattau_{s}}}{\hatcV}-\frac{2\hat{a}_{s}\left\lvert A_{s}W_{\text{GVW}}\right\rvert e^{-\hat{a}_{s}\hattau_{s}}\hattau_{s}}{\hatcV^{2}}+\frac{3\left\lvert W_{\text{GVW}}\right\rvert^{2}\hat{\xi}}{8\hatcV^{3}}\right]\, .
\end{align}
We can now stabilize both the blowup modulus and the volume by finding a local minimum of the potential where $\frac{\partial V}{\partial\hattau_{s}}=\frac{\partial V}{\partial\hatcV}=0$. Following the work of \cite{bbcq} and taking $\hat{a}_{s}\hattau_{s}\sim\text{ln }\hatcV\gg 1$ in order to cut off higher instanton corrections, one obtains the simple result
\begin{align}\label{eq:taumin}
\left\langle\hattau_{s}\right\rangle\simeq\left(\frac{3c\hat{\xi}}{16}\right)^{2/3}\hspace{1cm}\text{and}\hspace{1cm}\left\langle\hatcV\right\rangle\simeq\frac{\left\lvert W_{\text{GVW}}\right\rvert}{2c\hat{a}_{s}\left\lvert A_{s}\right\rvert}\sqrt{\hattau_{s}}e^{\hat{a}_{s}\hattau_{s}}\, .
\end{align}
Finally, we convert back to the string frame using the transformations $\hatcV=g_{s}^{-3/2}\cV$, $\hattau_{i}=g_{s}^{-1}\tau_{i}$, $\hat{a}_{i}=g_{s}a_{i}$, and $\hat{\xi}=g_{s}^{-3/2}\xi =-\frac{\chi (X)\zeta (3)}{2g_{s}^{3/2}}$. We find that\footnote{For a small number $h^{1,1}(X)$ of K\"ahler moduli relative to complex structure moduli $h^{2,1}(X)$, the Euler number will have a negative value $\chi (X)=2\left(h^{1,1}(X)-h^{2,1}(X)\right)<0$. Thus, the volume $\cV >0$.}
\begin{align}\label{eq:volmin}
\left\langle\tau_{s}\right\rangle&\simeq\frac{1}{4}\left(\frac{3c\chi (X)\zeta (3)}{4}\right)^{2/3}\, ,\\
\notag\\
\left\langle\cV\right\rangle&\simeq\frac{\left\lvert W_{\text{GVW}}\right\rvert}{2ca_{s}\left\lvert A_{s}\right\rvert}\sqrt{\tau_{s}}e^{a_{s}\tau_{s}}\, .
\end{align}

\end{appendices}







\end{document}